\documentclass[11pt]{article}
\usepackage{a4}
\usepackage{times}
\usepackage{amssymb}

\def\beq{\begin{equation}}
\def\bea{\begin{eqnarray}}
\def\eeq{\end{equation}}
\def\eea{\end{eqnarray}}
\def\R{$\mathcal{R}$}
\def\Z{$\mathbb{Z}$}
\def\SM{Standard Model}

\begin{document}
\input{epsf.sty}
\title{\bf  Towards the supersymmetric standard model from intersecting D6-branes on the $\mathbb{Z} _6'$ orientifold }
\maketitle
\vglue 0.35cm
\begin{center}
\author{\bf David Bailin \footnote
{D.Bailin@sussex.ac.uk} \&  Alex Love \\}
\vglue 0.2cm
	{\it  Department of Physics \& Astronomy, University of Sussex\\}
{\it Brighton BN1 9QH, U.K. \\}
\baselineskip=12pt
\end{center}
\vglue 2.5cm
\begin{abstract}
We construct  ${\mathcal N}=1$ supersymmetric fractional branes on the \Z$_6'$ orientifold. 
Intersecting stacks of such  branes are needed to build a supersymmetric standard model.
   If $a,b$  are the   stacks 
  that generate the  $SU(3)_c$ and $SU(2)_L$ gauge particles,  
 then,  in order to obtain {\em just} the chiral spectrum of the (supersymmetric)
  standard model (with non-zero Yukawa couplings to the Higgs mutiplets),
   it is necessary that  the number of intersections $a \circ b$ of the stacks $a$ and $b$, and 
  the number of intersections $a \circ b'$ of $a$ with the orientifold image $b'$ of $b$
   satisfy $(a \circ b,a \circ b')=\pm(2,1)$ or $\pm(1,2)$. 
It is also necessary that there is no matter in symmetric representations of the gauge group,
 and not too much matter in antisymmetric representations, on either stack.
We provide a number of examples having these properties. Different lattices give different solutions 
and different physics.
  
\end{abstract}

\newpage
\section{Introduction} \label{intro}
One of the main  phenomenological attractions of using D-branes  is that they permit a ``bottom-up'' approach 
to constructing the standard model from Type II string theory. Open strings that begin and end on a stack $a$ of $N_a$ D-branes 
generate the gauge bosons of 
a (supersymmetric) $U(N_a)$ gauge theory living in the world volume of the D-branes. 
In the original bottom-up models \cite{Aldazabal:2000sa,Bailin:2001ia,Berenstein:2001nk,Alday:2002uc}
 a stack of D3-branes is placed at an orbifold $T^6/\mathbb{Z}_N$ 
singularity and the standard model gauge group (possibly augmented by additional $U(1)$ factors)
 is obtained by choosing a suitable embedding $\gamma _{\theta}$  of the action
of the generator $\theta$ of the orbifold point group $\mathbb{Z}_N$ on the Chan-Paton indices of the D3-branes.
Besides the gauge bosons, fermionic matter also survives the 
orbifold projection. So long as only D3-branes are retained, the fermion spectrum generally makes 
the non-abelian gauge symmetries anomalous, reflecting the fact that  a general collection of D3-branes has uncancelled 
Ramond-Ramond (RR) tadpoles. The required cancellation is achieved by introducing D7-branes, which generate further gauge symmetries,
and additional fermions. When all tadpoles are cancelled, so are the gauge anomalies. 
However, we showed in an earlier paper \cite{Bailin:2004fh} that all such models, whether utilising fixed points on an orbifold 
or an orientifold, 
have electroweak Higgs content that is non-minimal, both for the (non-supersymmetric) Standard Model or 
its supersymmetric extension, the MSSM. As a consequence  there is 
a generic flavour changing neutral current (FCNC) problem in such models, and we conclude that such models are not realistic.
(See, however, \cite{Escudero:2005ku},
which argues that a supersymmetric, standard-like model with three Higgs doublets, derived from compactifying 
the $E_8 \otimes E_8$ heterotic string on a \Z$_3$ orbifold, {\em can} circumvent the FCNC problem without an 
excessively heavy Higgs sector.)

An alternative approach that also uses D-branes is  ``intersecting brane'' model building \cite{Lust:2004ks}. 
In these models one starts with 
 two stacks, $a$ and $b$ with $N_a=3$ 
and $N_b=2$, of  D4-, D5- or D6-branes wrapping the three large spatial 
dimensions plus respectively 1-, 2- and 3-cycles of the six-dimensional  internal space (typically a torus $T^6$ 
or a Calabi-Yau 3-fold) on which the theory is compactified.
 These generate  the gauge group $U(3) \times U(2) \ni SU(3) _c \times SU(2)_L$, and  the non-abelian component of the standard model gauge group
is immediately assured.
   Further, (four-dimensional) fermions in bifundamental representations 
$({\bf N} _a, \overline{\bf N}_b)= ({\bf 3}, \overline{\bf 2})$ 
of the gauge group can arise at the multiple intersections of the two stacks. 
These are precisely the representations needed for the quark doublets $Q_L$ of the Standard Model. For D4- and D5-branes,
 to get {\em chiral} fermions the stacks must be at a singular point of the transverse space. 
 In general, intersecting branes yield a non-supersymmetric spectrum, so that, to avoid the hierarchy problem, the string scale associated 
 with such models must be low, no more than a few TeV. Then, the high energy (Planck)  scale associated with gravitation 
does not emerge naturally. Nevertheless, it seems that these problems can be surmounted \cite{Blumenhagen:2002vp,Uranga:2002pg}, and indeed an 
attractive model having just the spectrum of the Standard Model has been constructed \cite{Ibanez:2001nd}. It uses D6-branes that wrap 3-cycles 
of an orientifold $T^6/\Omega$, where $\Omega$ is the world-sheet parity operator. The advantage and, indeed, the necessity of using 
an orientifold stems from the fact that for every stack $a,b, ...$ there is an orientifold image $a',b', ...$. 
At intersections of $a$ and $b$ there are chiral fermions 
in the $({\bf 3}, \overline{\bf 2})$ representation of $U(3) \times U(2)$, where the ${\bf 3}$ has charge $Q_a=+1$ with respect to the 
$U(1)_a$ in $U(3)=SU(3)_c \times U(1)_a$, and the $\overline{\bf 2}$ has charge $Q_b=-1$ with respect to the 
$U(1)_b$ in $U(2)=SU(2)_L \times U(1)_b$.  However, at intersections of $a$ and $b'$ there are chiral fermions 
in the $({\bf 3},{\bf 2})$ representation, where  the ${\bf 2}$ has $U(1)_b$ charge $Q_b=+1$. 
In the model of \cite{Ibanez:2001nd}, the number of intersections $a \circ b$ of the stack $a$ with $b$ is 2, 
and the number of intersections $a \circ b'$ of the stack $a$ with $b'$ is 1. Thus, as required for the \SM , there are 3 quark doublets.
 These have  net $U(1)_a$ charge $Q_a=6$, and net $U(1)_b$ charge $Q_b=-3$. Tadpole cancellation requires that overall both charges,
sum to zero, so further fermions are essential, and indeed required by the \SM. 6 quark-singlet states $u^c_L$ and $d^c_L$ 
belonging to the $({\bf 1}, \overline{\bf 3})$ representation of  $U(1) \times U(3)$, having 
a total of $Q_a=-6$ are sufficient to ensure overall cancellation of $Q_a$, and these arise from the intersections of $a$ with other 
stacks $c,d,...$ having just a single D6-brane. Similarly, 3 lepton doublets $L$, belonging to the $({\bf 2}, \overline{\bf 1})$
 representation of  $U(2) \times U(1)$, having 
a total $U(1)_b$ charge of $Q_b=3$, are sufficient to ensure overall cancellation of $Q_b$, 
and these arise from the intersections of $b$ with other 
stacks having just a single D6-brane. In contrast, had we not used an orientifold, the requirement of 3 quark doublets would 
necessitate having the number of intersections $a \circ b=3$. This makes no difference to the charge $Q_a=6$ carried by the quark doublets,   
but instead the $U(1)_b$ charge carried by the quark doublets is $Q_b=-9$, which cannot be cancelled by just 3 lepton doublets $L$. 
Consequently, additional vector-like fermions are unavoidable unless the orientifold projection is available.
This is why the orientifold is essential if we are to get just the matter content of the \SM \ or of the MSSM. 

 Actually, an orientifold  can allow the standard-model spectrum without vector-like matter even 
when $a \circ b=3$ and $a \circ b'=0$ \cite{Blumenhagen:2001te}. 
This is because in orientifold models it is also possible to get chiral matter in the symmetric and/or antisymmetric representation 
of the relevant gauge group from open strings stretched between a stack and its orientifold image. Both representations have charge $Q=2$ 
with respect to the relevant $U(1)$. The antisymmetric (singlet) representation of $U(2)$ can describe a lepton single state $\ell ^c_L$, and 
3 copies contribute $Q_b=6$ units of $U(1)_b$ charge. If there are also 3 lepton doublets $L$ belonging to the bifundamental representation 
$({\bf 2}, \overline{\bf 1})$
 representation of  $U(2) \times U(1)$, each contributing $Q_b=1$ as above, then the total contribution is $Q_b=9$ which {\bf can} 
 be cancelled by 3 quark doublets $Q_L$ in the $({\bf 3}, \overline{\bf 2})$ representation of $U(3) \times U(2)$. Thus, as asserted, 
 orientifold models can allow just
 the standard-model spectrum  even when $(a \circ b ,a \circ b')=(3,0)$.

Despite the attractiveness of the model in \cite{Ibanez:2001nd}, there remain serious problems in the absence of supersymmetry.
 A generic feature  of intersecting brane models 
is that flavour changing neutral currents 
are generated by four-fermion operators induced by string instantons \cite{Abel:2003yh}. The severe experimental limits on these processes 
require that the string scale is rather high, of order $10^4$ TeV. This makes the fine tuning problem very severe, and the viability 
of such models highly questionable. Further, in  non-supersymmetric theories, such as these, the cancellation of RR tadpoles does not ensure 
Neveu Schwarz-Neveu Schwarz (NSNS) tadpole cancellation. NSNS tadpoles are simply the first
 derivative of the scalar potential with respect to the scalar fields, specifically the complex structure  and K\"ahler moduli 
 and the dilaton. A non-vanishing derivative of the scalar potential signifies that  
 such scalar fields are not even solutions of the equations of motion. 
 Thus a particular consequence of the non-cancellation is that the complex structure moduli are unstable \cite{Blumenhagen:2001mb}. 
 One way to stabilise these moduli 
  is for the 
D-branes to wrap an orbifold $T^6/P$ rather than a torus $T^6$. 
The FCNC problem can be solved and the complex structure moduli stabilised when the theory is supersymmetric. 
First, a supersymmetric theory is not obliged to have the low string scale that led to problematic FCNCs  induced by string instantons. 
Second, 
 in a supersymmetric theory, RR tadpole cancellation ensures cancellation 
of the NSNS tadpoles \cite{Cvetic:2001tj,Cvetic:2001nr}.
An orientifold is then constructed by quotienting the orbifold with the world-sheet parity operator $\Omega$.
(
As explained above, an orientifold is necessary to allow the possibility of obtaining just the spectrum of the supersymmetric standard model.)

Several attempts to construct the MSSM using D6-branes
 and a \Z$_4$, \Z$_4 \times$\Z$_2$ or \Z$_6$ orientifold have been made \cite{Blumenhagen:2002gw, Honecker:2003vq,
Honecker:2004np,
Honecker:2004kb}. The most successful attempt to date is the last of these \cite{Honecker:2004kb, Ott:2005sa},
 which uses D6-branes intersecting on  a \Z$_6$ orientifold
 to construct an $\mathcal{N}=1$ supersymmetric standard-like model using 5 stacks of branes. 
We shall not discuss this beautiful model in any detail except to note that the intersection numbers for the stacks $a$, 
which generates the $SU(3)_c$ group, and $b$, which generates the $SU(2)_L$, are  $(|a \circ b|,|a \circ b'|)=(0,3)$.
 In this case it is impossible to obtain lepton singlet states $\ell ^c_L$ as antisymmetric representations of $U(2)$. 
Further, it was shown, quite 
generally, that it is impossible to find stacks $a$ and $b$ such that $(|a \circ b|,|a \circ b'|)=(2,1)$ or $(1,2)$. 
 Thus, as 
explained above, it is impossible to obtain exactly the (supersymmetric) \SM \ spectrum. 

The question then arises as to whether the use of a different orientifold could circumvent this problem. In this paper we 
address this question for the \Z$_6'$ orientifold. We do not attempt to construct a standard(-like) MSSM. Instead, we merely see 
whether there are any stacks $a,b$ that simultaneously satisfy the supersymmetry constraints, the absence of chiral matter in symmetric 
 representations of the gauge groups (see below), 
 which have not too much chiral matter in antisymmetric representations of the gauge groups (see below),
  and which have $(|a \circ b|,|a \circ b'|)=(2,1)$ or $(1,2)$.
   We do not pursue the alternative that $(|a \circ b|,|a \circ b'|)=(3,0)$ or $(0,3)$, since such models, with 3 lepton singlet states 
  arising on the $U(2)$ stack, do not have the  standard-model couplings of these states to the Higgs multiplet.  
 With $(N_a,N_b)=(3,2)$ we explained above why $(a \circ b, a \circ b')=(2,1)$ is sufficient to ensure that no vector-like matter is 
 necessary to ensure that the net $U(1)_b$ charge $Q_b$ is zero, and the same is obviously the true if $(a \circ b,  a \circ b')=(1,2)$;
  it amounts to interchanging $b$ and $b'$. Intersection numbers $(a \circ b , a \circ b')=(-2,-1)$ or $(-1,-2)$ are equally acceptable, 
  since negative intersection numbers correspond to opposite chiralities.  Thus $(a \circ b , a \circ b')=(\underline{\pm2,\pm1})$, where 
  underlining signifies any permutation, is sufficient. 
  For calculational purposes it is convenient to let {\em either} stack $a$ or $b$ generate the $SU(3)_c$ gauge group, so that
    $(N_a,N_b)=(\underline{3,2})$.   Interchanging $a$ and $b$ gives $(b \circ a , b \circ a')=(-a \circ b , a \circ b')$.
    Thus the intersection numbers are generally required to
     satisfy $(|a \circ b|,|a \circ b'|)=(\underline{2,1})$. 
      If $a \circ b $ and $ a \circ b'$ have the same sign, then $N_a=3$ and $N_b=2$; 
     otherwise $N_a=2$ and $N_b=3$.
In what follows we parallel quite closely the treatment \cite{Honecker:2004kb}
      of Honecker \& Ott for the \Z$_6$ orientifold. 
  
\section{The \Z$_6'$ orbifold} \label{z6'}
We assume, as is customary, that the torus $T^6$ factorises into three 2-tori $T^2_1 \times T^2_2 \times T^2_3$. 
The three 2-tori $T^2_k \ (k=1,2,3)$ are parametrised by three complex coordinates  $z_k$. 
 The action of the generator $\theta$ of the point group  $\mathbb{Z} ' _6$ on the   
coordinates $z_k $ is given by
\beq
\theta z_k = e^{2\pi i v_k} z_k
\eeq
where
\beq
 (v_1,v_2,v_3)
= \frac{1}{6} (1,2,-3)  \ {\rm for} \ \mathbb{Z} ' _6
 \label{z61vk}
\eeq
To calculate the number of independent bulk 3-cycles we need the Betti number  
 $b_3(T^6/\mathbb{Z}' _6)$ which is the dimension of the third homology group $H_3$ of the  space. 
 Because of the duality of the homology and cohomology groups we can as well compute the number of independent 
 invariant 3-forms. Then
 \beq
 b_3=b^3=\sum _{p+q=3} b^{p,q}
 \eeq
 where $b^{p,q}$ is the number of independent  invariant $p,q$ forms, {\it i.e.}with $p$ holomorphic  and $q$ 
 anti-holomorphic variables. For  $\mathbb{Z} ' _6$, the invariant forms are $dz_1 \wedge dz_2 \wedge dz_3$, 
   \ $dz_1 \wedge dz_2 \wedge d{\bar z}_3$ 
 and their complex conjugates,  so $b^{3,0}=1=b^{0,3}$ and $b^{2,1}=1=b^{1,2}$. Hence
,  the untwisted sector contributes
 \beq
b^{(0)}_3(T^6/\mathbb{Z}' _6)=4     \label{b3z6prime}
\eeq
 to the Betti number.

The point group action must be an automorphism of the lattice, so 
in $T^2_{1,2}$ we may take an $SU(3)$ lattice. Specifically we define the basis 1-cycles in $T^2_{1,2}$
 by $\pi _1$ and $\pi _2 \equiv e^{i\pi /3} \pi _1$ in $T^2_1$ and 
$\pi_3$ and 
$\pi _4 \equiv e^{i\pi /3} \pi _3$ in $T^2_2$. The orientation of $\pi _{1,3}$ relative to the real 
and imaginary axes of $z_{1,2}$ is arbitrary. 
 Since $\theta $ acts as a reflection in $T^2_3$ the lattice, with basis 1-cycles $\pi _5$ and $\pi _6$, 
 is arbitrary. The point group action on the basis 1-cycles is then 
\bea
\theta \pi _1 = \pi _2 \quad &{\rm and}& \quad \theta \pi _2 =\pi _2-\pi _1  \label{theta12} \\
\theta \pi _3  =\pi _4 -\pi _3 \quad &{\rm and}& \quad \theta \pi _4 =-\pi _3  \label{theta34} \\
\theta \pi _5=-\pi _5 \quad &{\rm and}& \quad \theta \pi _6 =-\pi _6 \label{thetaZ6}
\eea 
Now we  construct a basis of invariant 3-cycles.
With $\pi_{i,j,k} \equiv \pi _i \otimes \pi _j \otimes \pi _k$ where $i=1,2, \ j=3,4, \ k=5,6$ we define
the \Z$_6'$ invariant 3-cycle
\bea
\rho _1 & \equiv & (1+ \theta +\theta ^2 + \theta ^3 + \theta ^4 + \theta ^5) \pi _{1,3,5} \nonumber \\
&=& 2(1+ \theta +\theta ^2 ) \pi _{1,3,5} \nonumber \\
&=& 2(\pi _{1,3,5} + \pi _{2,3,5} +\pi _{1,4,5} - 2 \pi_{2,4,5}) \label{rho1}
\eea
In the same way
\bea
\rho _2 &\equiv &2(1+ \theta +\theta ^2 ) \pi _{2,3,5} \nonumber \\
 &=& 2(-\pi _{1,3,5} + 2\pi _{2,3,5} +2\pi _{1,4,5} -  \pi_{2,4,5})
\eea
and 
\bea
\rho _3 &\equiv &2(1+ \theta +\theta ^2 ) \pi _{2,4,5} \nonumber \\
 &=& 2(-2\pi _{1,3,5} + \pi _{2,3,5} +\pi _{1,4,5} +  \pi_{2,4,5}) \label{rho3} \\
 &=&\rho _2-\rho _1
\eea
Similarly, replacing $\pi _5 \rightarrow \pi _6$, we get
\bea 
\rho _4 &=&2(\pi _{1,3,6} + \pi _{2,3,6} +\pi _{1,4,6} - 2 \pi_{2,4,6})  \label{rho4}\\
\rho _5&=&2(-\pi _{1,3,6} + 2\pi _{2,3,6} +2\pi _{1,4,6} -  \pi_{2,4,6}) \\
\rho _6&=&2(-2\pi _{1,3,6} + \pi _{2,3,6} +\pi _{1,4,6} +  \pi_{2,4,6}) \label{rho6} \\
 &=&\rho _5-\rho _4
 \eea
Thus we can use the four cycles $\rho _1, \rho _3, \rho _4, \rho _6$ as the basis of the (untwisted) bulk 3-cycles, 
incidentally verifying (\ref{b3z6prime}). The most general  invariant bulk 3-cycle is 
\bea
2(1+ \theta +\theta ^2 ) \left[(n_1 \pi _1+m_1 \pi _2) \otimes (n_2 \pi _3+m_2 \pi _4) \otimes (n_3 \pi _5+m_3 \pi _6) \right] \nonumber \\
=A_1\rho _1+ A_3 \rho_3 +A_4\rho _4+ A_6 \rho_6 \label{genbulk}
\eea
where $(n_k, m_k) \ (k=1,2,3)$ are the (coprime) wrapping numbers on the $k$th torus, and
\bea 
A_1= (n_1n_2+n_1m_2+ m_1n_2)n_3  \label{A1}\\
A_3= (m_1m_2+n_1m_2+ m_1n_2)n_3 \\
A_4= (n_1n_2+n_1m_2+ m_1n_2)m_3 \\
A_6= (m_1m_2+n_1m_2+ m_1n_2)m_3 \label{A6}
\eea 
The intersection number is defined as
\beq
\Pi _a \circ \Pi _b \equiv \frac{1}{6} \left( \sum _{i=0}^5 \theta ^i \pi _a \right) \circ \left( \sum _{j=0}^5 \theta ^j \pi _b \right)
\label{PiaoPib}
\eeq
which gives 
\bea
\rho _1 \circ \rho _3=&0&= \rho _4 \circ \rho _6 \\
\rho _1 \circ \rho _4=-4, &\quad & \rho _1 \circ \rho _6=2 \\
\rho _3 \circ \rho _4=2, &\quad & \rho _3 \circ \rho _6=-4 
\eea
In general, if $\pi _a$ has wrapping numbers $(n^a_k, m^a_k)$ $(k=1,2,3)$, and $\pi _b$ has wrapping numbers $(n^b_k, m^b_k)$, then 
the intersection number of the orbifold-invariant 3-cycles $\Pi _a $ and $\Pi _b$  generated from $\pi _a$ and $\pi _b$ is
\bea
\Pi _a \circ \Pi _b = -4(A^a_1A^b_4&-&A^a_4A^b_1)+ 2(A^a_1A^b_6-A^a_6A^b_1) +2(A^a_3A^b_4-A^a_4A^b_3) - \nonumber \\
 &-&4(A^a_3A^b_6-A^a_6A^b_3) \qquad \equiv F(A^a_p,A^b_p)      \label{pia0pib}
\eea
which is always even. Here $A^a_p \ (p=1,3,4,6)$ relate to $\Pi _a$ and are given by (\ref{A1}) -  (\ref{A6})
with the wrapping numbers $(n^a_k, m^a_k)$, and similarly for  the $A^b_p$ which relate  to $\Pi _b$.

Besides the (untwisted) bulk 3-cycles discussed above, there are also exceptional 3-cycles associated with (some of) the 
twisted sectors of the orbifold. They arise in twisted sectors in which there is a fixed torus, 
and consist of a collapsed 2-cycle at a fixed point times a 1-cycle in the invariant plane.
 For the \Z$_6'$ orbifold the 
$\theta ^2$  and  $\theta ^4$ twisted sectors leave $T^2_3$ invariant, and the $\theta ^3$ twisted sector leaves $T^2_2$ invariant. 

In the $\theta ^2$ and $\theta ^4$ twisted sectors, there is a \Z$_3$ symmetry which has nine fixed points at
\beq
 e_{i,j} \equiv \frac{i}{3}(e_1+e_2) \otimes \frac{j}{3}(e_3+e_4)
 \eeq
 where $i,j=0,1,2  \bmod{3}$,  $e_{1}$ and $e_2 \equiv e^{i\pi /3}e_1$ are the basis lattice vectors in $T^2_1$, and  
  $e_{3}$ and $e_4 \equiv e^{i\pi /3}e_3$ are the basis lattice vectors in $T^2_2$. The   $\mathbb{Z}' _6$ generator $\theta$ acts on the fixed points as
  \bea
  \theta  \ \frac{i}{3}(e_1+e_2)=-\frac{i}{3}(e_1+e_2) \\
  \theta  \ \frac{j}{3}(e_3+e_4)=\frac{j}{3}(e_3+e_4)
  \eea
  Exceptional cycles in these sectors have the form $e_{i,j} \otimes (n_3 \pi _5+m_3 \pi _6)$. The action of the point group is given by
  \bea
  \theta \  e_{i,j} \otimes (n_3 \pi _5+m_3 \pi _6) &\equiv & e_{\theta i, \theta j} \otimes \theta(n_3 \pi _5+m_3 \pi _6) \nonumber \\
  &=&  e_{-i, j} \otimes (-n_3 \pi _5-m_3 \pi _6) 
  \eea
  Thus there are six invariant 3-cycles in each $\theta ^n  \ (n=2,4)$ twisted sector. As a basis we may choose
  \beq 
  \eta^{(n)} _j \equiv (e^{(n)}_{1,j}-e^{(n)}_{2,j}) \otimes \pi_5, \quad \tilde{\eta}^{(n)} _j \equiv (e^{(n)}_{1,j}-e^{(n)}_{2,j}) \otimes \pi _6 \quad (j=0,1,2) 
  \label{etaj}
  \eeq
  where $e^{(n)}_{i,j}$ is the collapsed 2-cycle at the fixed point $e_{i,j}$ in the $\theta ^n$ twisted sector.
  The intersection numbers  of these 2-cycles may be computed by blowing up the \Z$_3$ singularities \cite{Blumenhagen:2002wn}. 
  This yields an intersection matrix which is the Cartan matrix ${\bf A}_{(2)}$ of the Lie algebra of $A_{(2)} \cong SU(3)$. Then, 
  using  the intersection numbers of the 1-cycles on $T^2_3$, we find that
 
\beq
\eta ^{(m)}_j \circ \tilde{\eta}^{(n)}_k =-\delta _{jk} A_{(2)}^{mn}  \label{etaj0k}
\eeq 
where
\beq
{\bf A}_{(2)}=\left( \begin{array}{rr}
2 &-1 \\
-1& 2
\end{array} \right)
\eeq

Exceptional cycles also occur in the $\theta ^3$ sector. There is a $\mathbb{Z}_2$ symmetry 
acting in $T^2_1$ and $T^2_3$ and this has sixteen fixed points at
\beq
f_{i,j} \equiv \frac{1}{2}(\sigma _1 e_1+ \sigma _2e_2) \otimes \frac{1}{2}(\tau _1 e_5+ \tau _2e_6) \label{fij}
\eeq
where $\sigma _{1,2}, \tau_{1,2} =0,1 \pmod{2}$, and, using the notation of reference \cite{Honecker:2004kb},
 $i,j=1,4,5,6$ correspond to the pairs
$(\sigma _1,\sigma _2)$ or $(\tau _1,\tau_2)$
\beq
1 \sim (0,0), \quad 4 \sim (1,0), \quad 5 \sim (0,1), \quad 6 \sim (1,1) \label{1456}
\eeq 
The   $\mathbb{Z}' _6$ generator $\theta$ acts on the fixed points as
  \bea
  \theta  \ \frac{1}{2}(\sigma _1e_1+\sigma _2e_2)&=&\frac{1}{2}\left[-\sigma _2e_1+(\sigma _1+\sigma _2)e_2 \right] \\
  \theta  \ \frac{1}{2}(\tau _1e_5+\tau _2e_6)&=&-\frac{1}{2}(\tau _1e_5+\tau _2e_6)=\frac{1}{2}(\tau _1e_5+\tau _2e_6)
  \eea
  so 
  \bea
  \theta  \ f_{1,j}=f_{1,j}, \quad \theta  \ f_{4,j}=f_{5,j}  \label{f14j}\\
\theta  \ f_{5,j}=f_{6,j}, \quad \theta  \ f_{6,j}=f_{4,j} \label{f56j}
\eea 
The exceptional cycles are then $f_{i,j} \otimes (n_2 \pi _3+m_2\pi_4)$. 
Using (\ref{theta34}) we construct the orbifold invariant exceptional 3-cycles:
\bea
(1+\theta +\theta ^2) f_{1,j} \otimes \pi _3 &=&0 =(1+\theta +\theta ^2) f_{1,j} \otimes \pi _4\\
(1+\theta +\theta ^2) f_{6,j} \otimes \pi _3 &=& (f_{6,j}-f_{4,j}) \otimes 
\pi _3+ (f_{4,j}-f_{5,j}) \otimes \pi _4 \equiv \epsilon _j  \label{epsj}\\
(1+\theta +\theta ^2) f_{4,j} \otimes \pi _3 & =&(f_{4,j}-f_{5,j}) \otimes \pi _3+ (f_{5,j}-f_{6,j}) \otimes \pi _4 
 \equiv \tilde{\epsilon} _j \label{epstilj}\\
(1+\theta +\theta ^2) f_{5,j} \otimes \pi _3 & =&(f_{5,j}-f_{6,j}) \otimes 
\pi _3+ (f_{6,j}-f_{4,j}) \otimes \pi _4  \nonumber \\
&  =& -\epsilon _j - \tilde{\epsilon} _j 
\eea
It is easy to see that $f_{i,j} \otimes \pi _4$ generate the same orbifold invariants, 
so there are just eight independent orbifold-invariant 
exceptional 3-cycles $\epsilon _j$ and $\tilde{\epsilon} _j \ (j=1,4,5,6)$.
The non-zero intersection numbers for the  invariant combinations (\ref{epsj}) and (\ref{epstilj}) are given by
\beq
\epsilon _j \circ \tilde{\epsilon} _k=-2 \delta _{jk} \label{eps0jk}
\eeq
using the self-intersection number of  $-2$ for an exceptional 2-cycle at the $\mathbb{Z}_2$ fixed point 
 $f_{4,j}$ and the intersection numbers of the 
1-cycles on $T^2_2$. The relation between fixed points and exceptional cycles is shown in Table \ref{FPex}. 
\begin{table}
 \begin{center}
\begin{tabular}{||c|c||} \hline \hline
Fixed point $\otimes$ 1-cycle & Exceptional 3-cycle \\ \hline \hline
$f_{1,j} \otimes (n_2 \pi _3 +m_2 \pi _4)$ & 0 \\ \hline
$f_{4,j} \otimes (n_2 \pi _3 +m_2 \pi _4)$ & $m_2 \epsilon _j + (n_2+m_2) \tilde{\epsilon _j}$ \\ \hline
$f_{5,j} \otimes (n_2 \pi _3 +m_2 \pi _4)$ & $-(n_2+m_2) \epsilon _j-n_2 \tilde{\epsilon _j} $ \\ \hline
$f_{6,j} \otimes (n_2 \pi _3 +m_2 \pi _4)$ & $n_2 \epsilon _j-m_2 \tilde{\epsilon _j} $ \\ \hline \hline
\end{tabular}
\end{center} 
\caption{ \label{FPex} Relation between fixed points and exceptional 3-cycles.}
\end{table}
\section{The \Z$_6'$ \ orientifold} \label{z6'orientifold}
We have already noted that the use of an orientifold is necessary if we are to obtain just the spectrum of the MSSM. In fact, 
like the D3-planes in the bottom-up models, the
net cancellation of RR charge cannot be achieved using just D6-branes wrapping an orbifold. The use of an orientifold is
also necessary for the cancellation of the RR charge. The embedding $\mathcal{R}$ of the world-sheet parity operator $\Omega$ 
by an anti-holomorphic involution induces O6-planes that carry negative RR charge, and it this that enables the necessary cancellation.
The action of  \R \  on the three complex coordinates $z_k \ (k=1,2,3)$ is
\beq
\mathcal{R}z_k=e^{i\phi _k}\overline{z} _k
\eeq
where $\phi _k$ is an arbitrary phase that we may choose to be zero. Thus, \R \ acts as complex conjugation,  
and we require that this too is an automorphism of the lattice. 
This fixes the orientation of the basis 1-cycles in each torus relative to the 
Re$z_k$ axis. It requires them to be in one of two configurations {\bf A} or {\bf B}. 

\begin{figure*}[h]
\center{
\leavevmode
\epsfbox{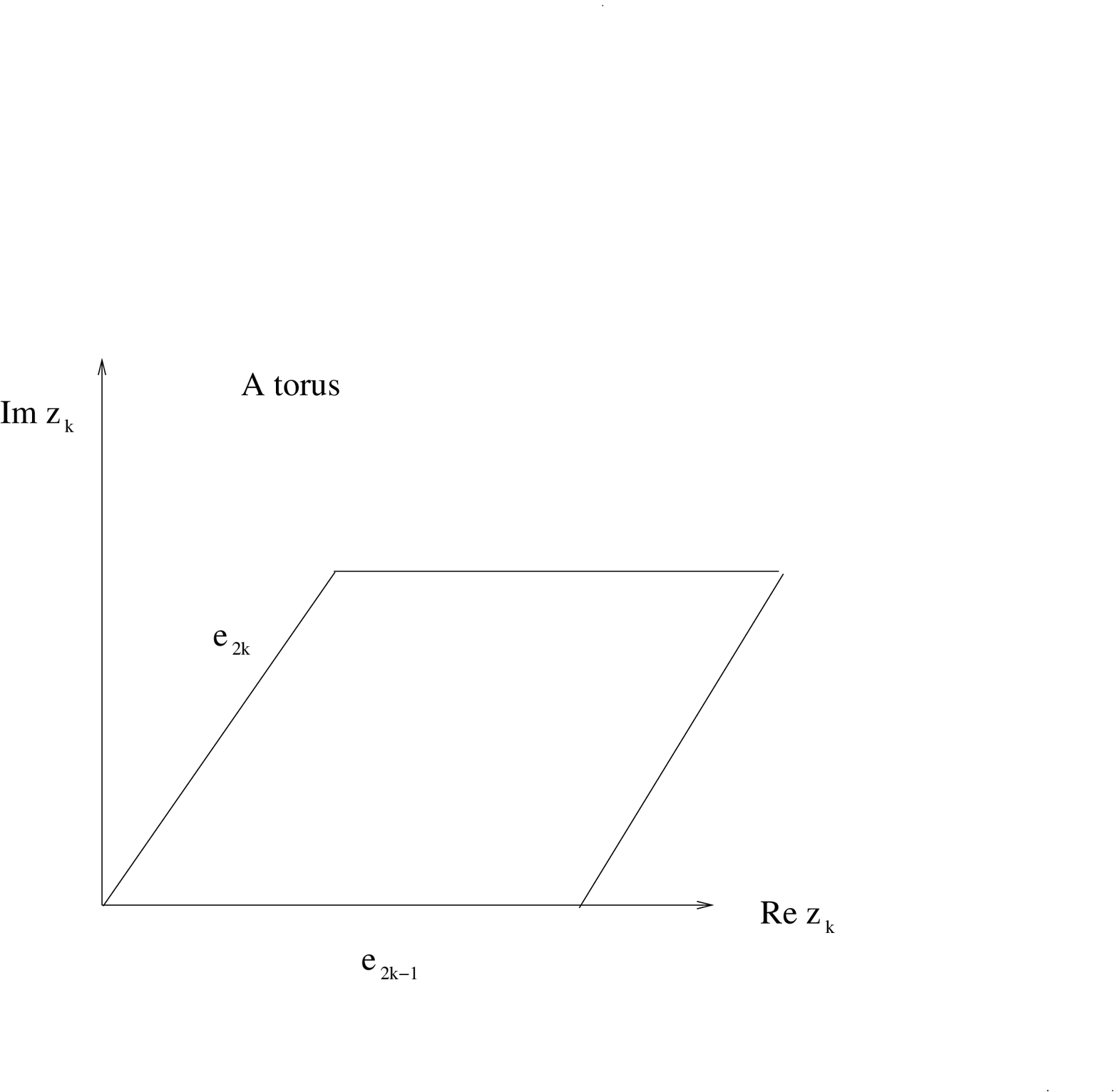}
\caption{The fundamental tori $T^2_k, \ k=1,2$ in the {\bf A} configuration.}
\label{TkAfig}
}
\end{figure*}

\begin{figure*}[h]
\center{
\leavevmode
\epsfbox{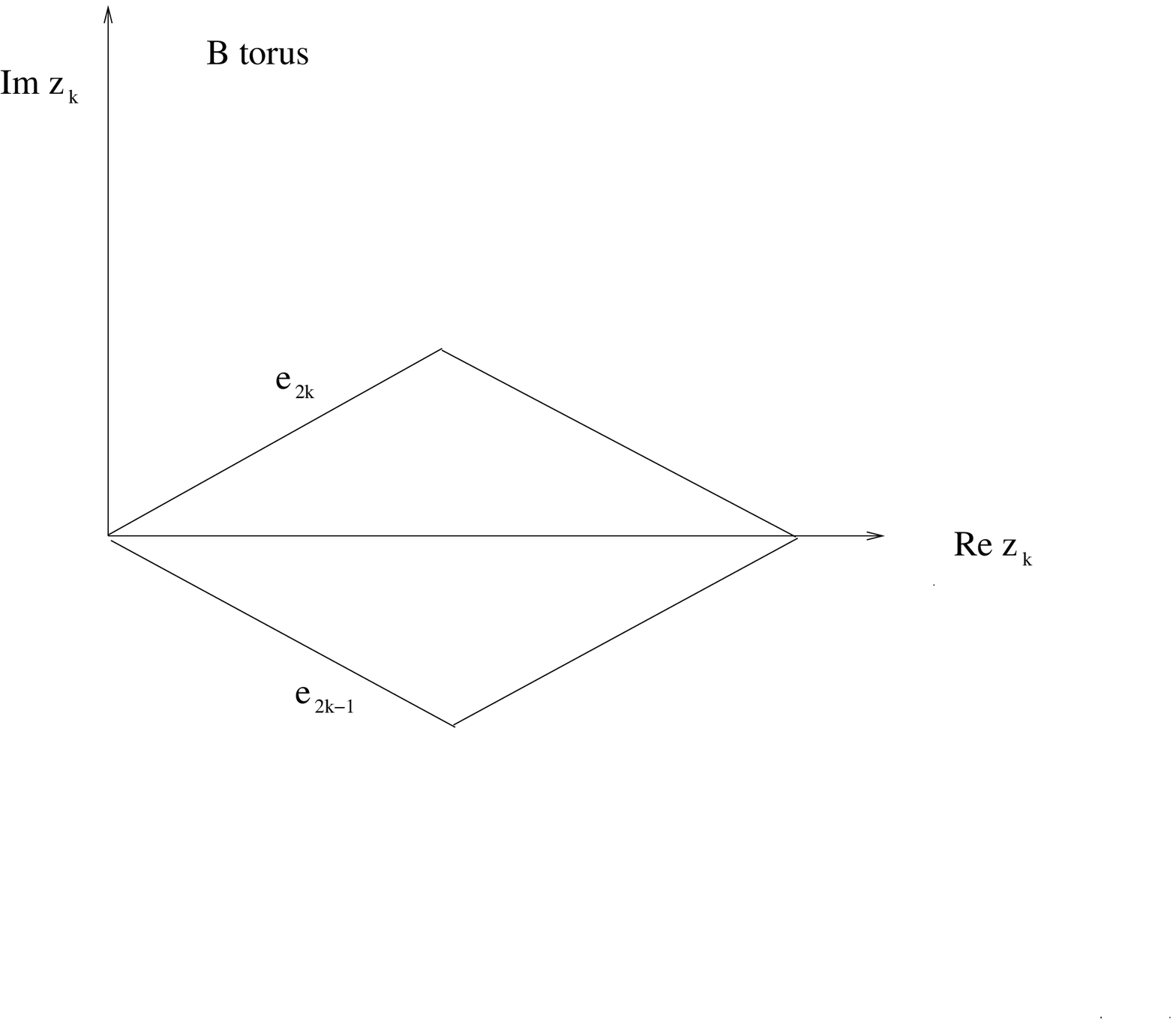}
\caption{The fundamental tori $T^2_k, \ k=1,2$ in the {\bf B} configuration.}
\label{TkBfig}
}
\end{figure*}

\begin{figure*}[h]
\center{
\leavevmode
\epsfbox{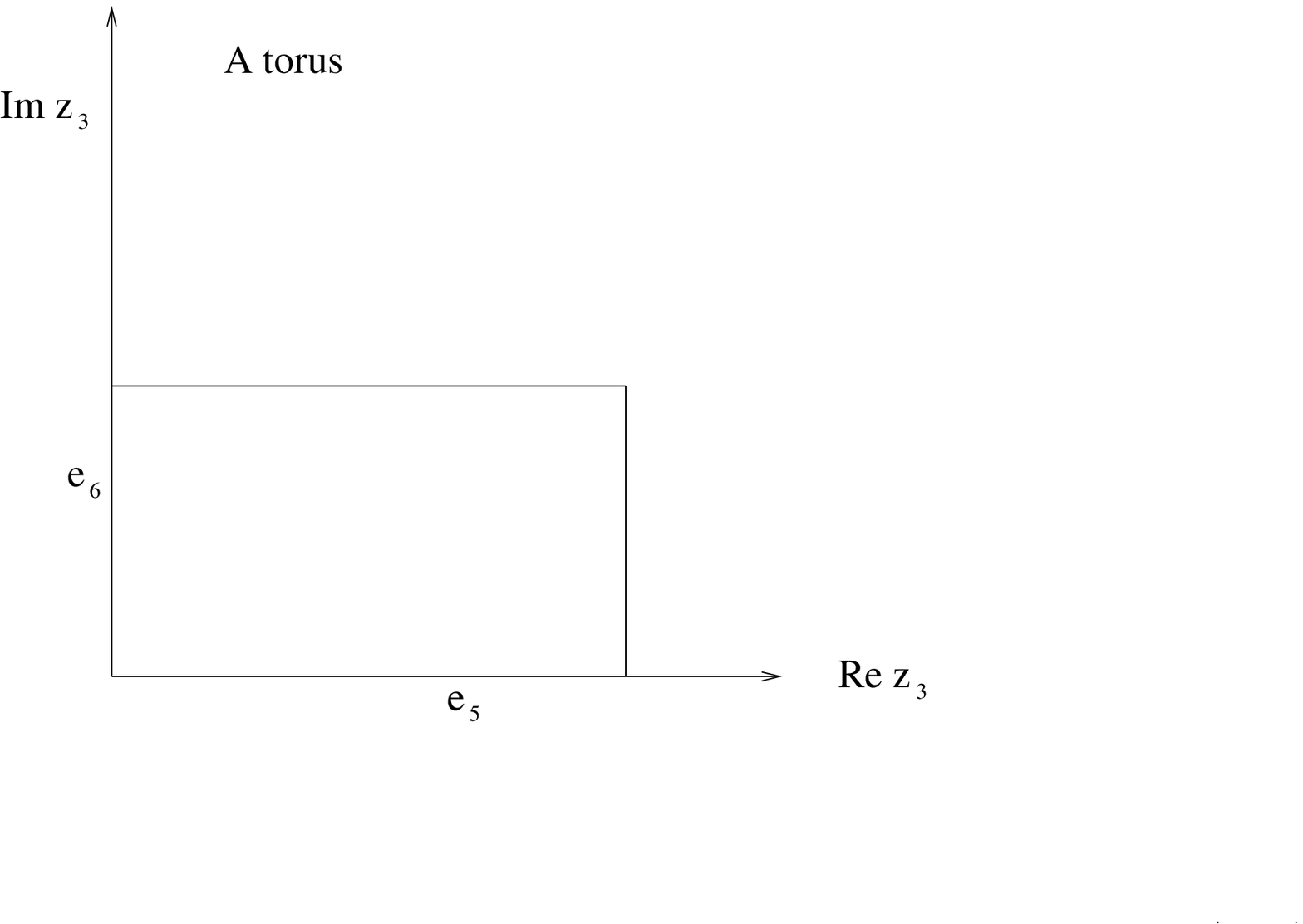}
\caption{The fundamental torus $T^2_3,$ in the {\bf A} configuration.}
\label{T3Afig}
}
\end{figure*}

\begin{figure*}[h]
\center{
\leavevmode
\epsfbox{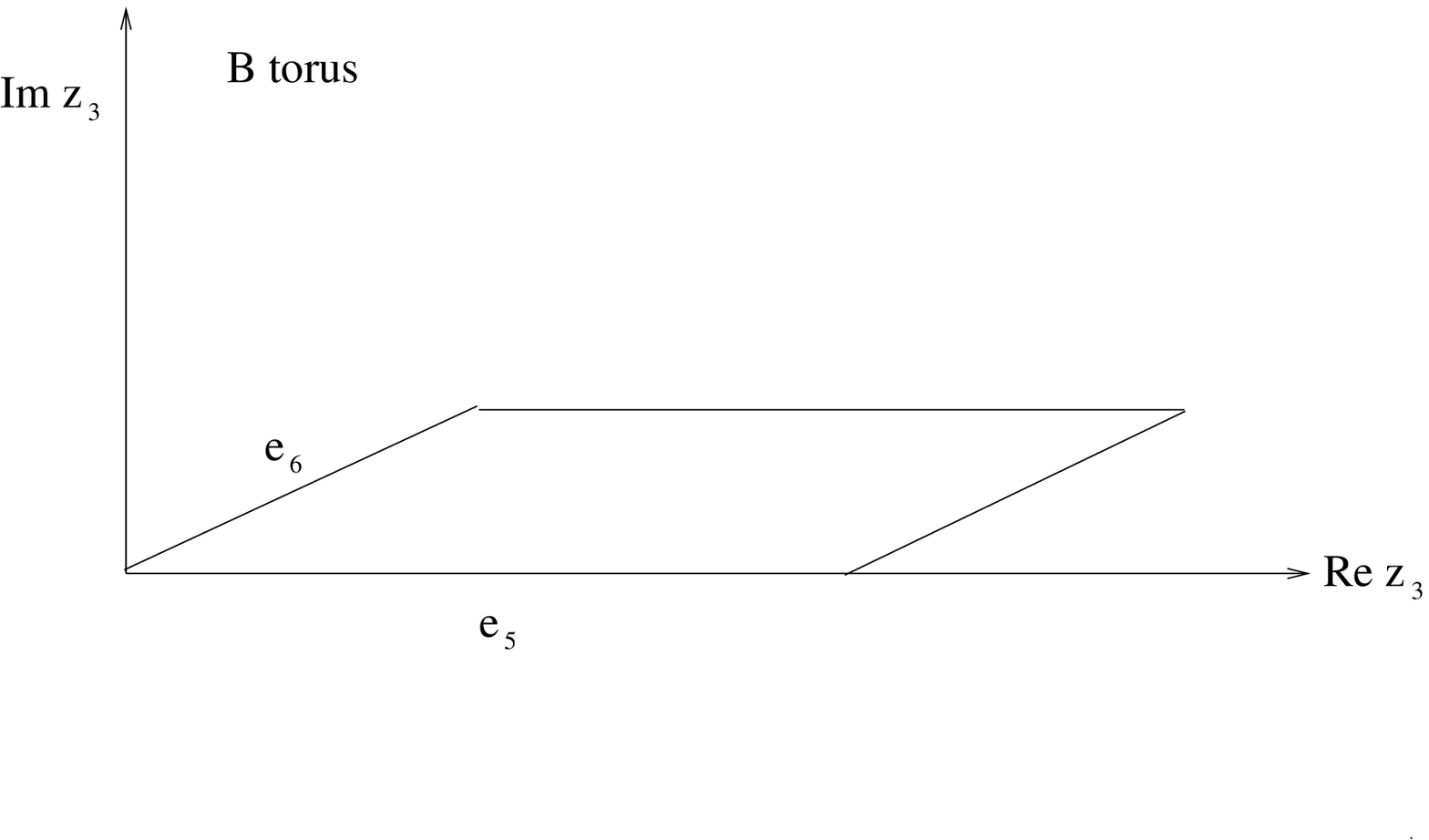}
\caption{The fundamental torus $T^2_3$ in the {\bf B} configuration.}
\label{T3Bfig}
}
\end{figure*}

For $k=1,2$
\bea
{\rm  {\bf{A}}}: \ & {\cal R}\pi _{2k-1}= \pi _{2k-1}, \quad &  {\cal R}\pi _{2k}= \pi _{2k-1}- \pi _{2k} \label{RA1} \\
{\rm  {\bf{B}}}: \  & {\cal R}\pi _{2k-1}= \pi _{2k}, \quad &  {\cal R}\pi _{2k}= \pi _{2k-1} \label{RB1}
\eea
In either case, the complex structure $U _k$ is given by
\beq
U _k \equiv \frac{\pi _{2k}}{\pi _{2k-1}} = e^{i \pi /3}= \frac{1}{2} +i \frac{\sqrt{3}}{2} \label{tau12}
\eeq 
The fundamental tori in the two configurations are shown in Figs \ref{TkAfig}, \ref{TkBfig}; $\pi _{2k-1,2k}$ wraps $e_{2k-1,2k}$ respectively.
\epsfysize=8 cm
\epsfysize=8 cm
On $T^2_3$ we  have 
\bea
{\rm {\bf A}}: \ {\cal R}\pi _{5}=\pi _{5} \quad &{\rm and}& \quad {\cal R}\pi _{6}=-\pi _{6} \label{RA31} \\
{\rm {\bf B}}: \ {\cal R}\pi _{5}=\pi _{5} \quad &{\rm and}& \quad {\cal R}\pi _{6}=\pi _{5}-\pi _6 \label{RB31}
\eea
In both cases $\pi _5$ is real, and in the {\bf A} case $\pi _6$ is pure imaginary. Thus
\beq
U _3 ^{\bf A} = i \frac{R_6}{R_5}
\eeq
where $R_{5,6}$ are the length of the 1-cycles $\pi_{5,6}$.
In the {\bf B} case, 
$\pi _6 +\cal{R}\pi _6= \pi _5$, so that
\beq
U _3^{\bf B} = \frac{1}{2} + i\sqrt{\frac{R_6^2}{R_5^2}-\frac{1}{4}} \label{tau3B}
\eeq
Both are summarised in the formula
\beq
U_3= b_3 + i\sqrt{\frac{R_6^2}{R_5^2}-b_3^2} \label{tau3}
\eeq
where $b_3=0, \frac{1}{2}$ respectively  for the {\bf A, B} orientations. 
The fundamental tori in the two configurations are shown in Figs \ref{T3Afig}, \ref{T3Bfig}; $\pi _{5,6}$ wrap $e_{5,6}$ respectively,
\epsfysize=8 cm
\epsfysize=8 cm

 The ${\cal R}$-images of the four basis bulk 3-cycles $\rho _{1,3,4,6}$,
  defined in (\ref{rho1}),(\ref{rho3}),(\ref{rho4}) and (\ref{rho6}),
 on each of the lattices may now be calculated. They
 are given in Table \ref{Rrhoz61}. 
\begin{table}
 \begin{center}
\begin{tabular}{||c||c|c|c|c||} \hline \hline
Lattice& ${\cal R}\rho _1$ & ${\cal R}\rho _3$ & ${\cal R}\rho _4$ & ${\cal R}\rho _6$ \\ \hline \hline
{\bf AAA} &$\rho _1$ & $ -\rho _1 - \rho _3$ & $ - \rho _4$ & $\rho _4 +\rho _6$ \\ \hline
{\bf AAB} &$\rho _1$ & $ -\rho _1-\rho _3 $ & $\rho _1-\rho _4 $ & $-\rho _1 - \rho _3+\rho _4 +\rho _6$ \\ \hline
{\bf ABA} and {\bf BAA} &$\rho _1+\rho _3$ & $ - \rho _3$ & $-\rho _4 - \rho _6$ & $\rho _6$ \\ \hline
{\bf ABB} and {\bf BAB} &$\rho _1+\rho _3$ & $ -\rho _3 $ & $\rho _1 + \rho _3-\rho _4 - \rho _6$ & $ \rho _6- \rho _3$ \\ \hline
{\bf BBA} &$\rho _3$ & $ \rho _1 $ & $- \rho _6$ & $- \rho _4$ \\ \hline
{\bf BBB} &$\rho _3$ & $ \rho _1$ & $\rho _3-\rho _6$ & $  \rho _1-\rho _4$ \\ 
\hline \hline
\end{tabular}
\end{center} 
\caption{ \label{Rrhoz61} ${\cal R}$-images of the bulk 3-cycles.}
 \end{table}
The O6-planes that are invariant under ${\cal R}$ may then be identified. 
In each case there are two linearly independent ${\cal R}$-invariant combinations, 
which may be chosen to be those given in Table \ref{O6Z61planes}.
\begin{table}
 \begin{center}
\begin{tabular}{||c||c|c||} \hline \hline
{\bf AAA} & $\rho _1$ & $ \rho _4+2\rho _6$ \\ \hline 
{\bf AAB} & $\rho _1$ & $ -\rho _3+\rho _4 +2 \rho _6$ \\ \hline
{\bf ABA} and {\bf BAA} & $2\rho _1+\rho _3$ & $\rho _6$ \\ \hline
{\bf ABB} and {\bf BAB} & $2\rho _1 +\rho _3 $ & $-\rho _3 +2\rho _6$ \\ \hline
{\bf BBA} & $\rho _1 +\rho _3$ & $\rho _4-\rho _6$ \\ \hline
{\bf BBB} & $\rho _1 +\rho _3 $ & $\rho _1-\rho _3-2\rho _4+2\rho _6$ \\ \hline \hline
\end{tabular}
\end{center} 
\caption{ \label{O6Z61planes} ${\cal R}$-invariant bulk 3-cycles.}
 \end{table}
The cancellation of RR tadpoles requires that the overall homology class of the D6-branes and O6-planes vanishes:
\beq
\sum_a N_a (\Pi _a+\Pi _a')-4\Pi _{{\rm O}6}=0 \label{RRtad}
\eeq
where $N_a$ is the number of D6-branes in the stack $a$, $\Pi _a'$ is the orientifold dual of $\Pi _a$
\beq
\Pi _a' \equiv {\cal R}\Pi _a
\eeq
and $\Pi _{\rm O6}$ is the homology class of the O6-planes.
To determine the last of these we must consider the factorisable 3-cycles in detail. 
 The ${\cal R}$- and  $\theta {\cal R}$-invariant  cycles  on the three tori $T^2_k$ are as given in Table \ref{RthetainvZ61}.
\begin{table}
 \begin{center}
\begin{tabular}{||c|c||c|c|c||} \hline \hline
Lattice & Invariant & $T^2_1$ &$T^2_2$ &$T^2_3$ \\ \hline \hline
{\bf A} & $ {\cal R}$ & $\pi _1$ & $\pi _3$ &$\pi _5$ \\ 
&  $ \theta{\cal R}$ & $\pi _1+\pi _2$ & $\pi _3+\pi _4$ &$\pi _6$ \\ \hline 
{\bf B} & $ {\cal R}$ & $\pi _1+\pi _2$ & $\pi _3+\pi _4$ &$\pi _5$ \\ 
&  $ \theta{\cal R}$ & $\pi _2$ & $\pi _3-2\pi _4$ &$\pi _5 - 2\pi _6$ \\ \hline \hline
\end{tabular}
\end{center} 
\caption{ \label{RthetainvZ61} ${\cal R}$- and  $\theta {\cal R}$-invariant cycles.}
 \end{table}
\begin{table}
 \begin{center}
\begin{tabular}{||c|c|c|c||} \hline \hline
 Lattice & Invariant & $(n_1,m_1)(n_2,m_2)(n_3,m_3)$ &  3-cycle \\ \hline \hline 
 {\bf AAA} & ${\cal R}$ & $(1,0)(1,0)(1,0)$ & $\rho _1$ \\ 
  &$\theta{\cal R}$& $(1,1)(0,1)(0,1)$  & $\rho _4+2\rho _6$ \\ \hline
   {\bf AAB} & ${\cal R}$& $(1,0)(1,0)(1,0)$  & $\rho _1$ \\ 
   &$\theta{\cal R}$& $(1,1)(0,1)(1,-2)$  &$\rho _1+2\rho _3-2\rho _4-4\rho _6$ \\ \hline
{\bf ABA} &${\cal R}$ & $(1,0)(1,1)(1,0)$  & $2\rho _1+\rho _3$ \\ 
  &$\theta{\cal R}$& $(1,1)(1,-2)(0,1)$ &$-3\rho _6$ \\ \hline
{\bf BAA} &${\cal R}$ & $(1,1)(1,0)(1,0)$  & $2\rho _1+\rho _3$ \\ 
&$\theta{\cal R}$& $(0,1)(0,1)(0,1)$ &$\rho _6$ \\ \hline
{\bf ABB} &${\cal R}$ & $(1,0)(1,1)(1,0)$  & $2\rho _1 +\rho _3$ \\
  &$\theta{\cal R}$& $(1,1)(1,-2)(1,-2)$ &$-3\rho _3+6\rho _6$ \\  \hline
{\bf BAB} &${\cal R}$ & $(1,1)(1,0)(1,0)$  & $2\rho _1 +\rho _3$ \\
  &$\theta{\cal R}$& $(0,1)(0,1)(1,-2)$ &$\rho _3-2\rho _6$ \\  \hline
{\bf BBA} &${\cal R}$ & $(1,1)(1,1)(1,0)$  & $3\rho _1+3\rho _3$ \\
  &$\theta{\cal R}$ & $(0,1)(1,-2)(0,1)$ &$\rho _4-\rho _6$ \\  \hline
{\bf BBB} &${\cal R}$ & $(1,1)(1,1)(1,0)$ & $3\rho _1+3\rho _3$ \\
  &$\theta{\cal R}$ & $(0,1)(1,-2)(1,-2)$  &$\rho _1-\rho _3-2\rho _4+2\rho _6$ \\  \hline \hline
\end{tabular}
\end{center} 
\caption{ \label{pio6Z61} O6-planes of the $\mathbb{Z}' _6$ orientifold. The total homology class $\Pi_{\rm O6}$
is obtained by summing over the two orbits for each lattice.}
\end{table} 
These generate point-group invariant 3-cycles for each lattice, as given in Table \ref{pio6Z61}. 
The total homology class is obtained by summing the two contributions. Thus RR tadpole cancellation (\ref{RRtad}) requires that
\bea
{\rm \bf AAA}: &\quad& \sum _aN_a \left[ (2A_1^a-A_3^a) \rho _1+A_6^a(\rho _4+2\rho_6) \right]=4(\rho _1+\rho _4+2\rho _6) \label{RRbulkaaa} \\
{\rm \bf AAB}: &\quad& \sum _aN_a \left[ (2A_1^a-A_3^a+A_4^a-A_6^a) \rho _1-A_6^a(\rho _3-\rho _4-2\rho_6) \right]
=8(\rho _1+\rho _3 -\rho _4-2\rho _6) \nonumber \\
&& \\
{\rm \bf ABA}: &\quad& \sum _aN_a \left[A_1^a (2\rho _1+\rho_3)+(2A_6^a-A^a _4)\rho_6 \right]=4(2\rho _1+\rho _3-3\rho _6) \\
{\rm \bf BAA}: &\quad& \sum _aN_a \left[A_1^a (2\rho _1+\rho_3)+(2A_6^a-A^a _4)\rho_6 \right]=4(2\rho _1+\rho _3+\rho _6) \\
{\rm \bf ABB}: &\quad& \sum _aN_a \left[ (2A_1^a+A_4^a) \rho _1+(A_1^a+A_4^a-A_6^a)\rho _3+(2A_6^a-A_4^a)\rho_6) \right]=8(\rho _1-\rho _3+3\rho _6) \nonumber \\
&& \\
{\rm \bf BAB}: &\quad& \sum _aN_a \left[ (2A_1^a+A_4^a) \rho _1+(A_1^a+A_4^a-A_6^a)\rho _3+(2A_6^a-A_4^a)\rho_6) \right]=8(\rho _1+\rho _3-\rho _6) \nonumber \\
&& \\
{\rm \bf BBA}: &\quad& \sum _aN_a \left[(A_1^a+A_3^a)(\rho _1+\rho_3)+(A_4^a-A^a _6)(\rho _4-\rho_6) \right]=4(3\rho _1+3\rho _3+\rho _4+\rho _6) \\
{\rm \bf BBB}: &\quad& \sum _aN_a \left[(A_1^a+A_3^a+A_6^a)\rho _1+(A_1^a+A_3^a+A_4^a)\rho_3+(A_4^a-A^a _6)(\rho _4-\rho_6) \right] \nonumber \\
&& \qquad \qquad \qquad =8(2\rho _1+\rho _3-\rho _4+\rho _6) 
\label{RRbulkbbb}
\eea
where $A_p^a \ (p=1,3,4,6)$ are the bulk coefficients for the stack $a$. In each case we obtain two constraints, there being two independent 
$\mathcal{R}$-invariant combinations of the four basis bulk 3-cycles.
Note that, unlike in the \Z$_6$ case, the individual contributions from each stack $a$ do not generally wrap the O6-plane.

 The orientifold images of the exceptional cycles may also be computed. For the $\theta ^2$- and $\theta ^4$-sector exceptional cycles
 $\eta^{(n)} _j$ and $\tilde{\eta}^{(n)} _j$ with $n=2,4$ and $j=0,1,2 \bmod 3$, defined in (\ref{etaj}), they are given 
  in Table \ref{Repsij} on all eight lattices.
\begin{table}
 \begin{center}
\begin{tabular}{||c||c|c||} \hline \hline
Lattice & ${\cal R} \eta^{(n)} _{j}$ & ${\cal R} \tilde{\eta}^{(n)} _{j}$ \\ \hline \hline
{\bf AAA} & $-\eta^{(n)} _{-j}$ & $\tilde{\eta} ^{(n)}_{-j}$ \\ \hline
{\bf AAB} & $-\eta^{(n)} _{-j}$ & $-\eta^{(n)} _{-j} +\tilde{\eta}^{(n)} _{-j}$ \\ \hline
{\bf ABA} & $-\eta^{(n)} _{j}$ & $\tilde{\eta}^{(n)} _{j}$ \\ \hline 
{\bf BAA} & $\eta^{(n)} _{-j}$ & $-\tilde{\eta}^{(n)} _{-j}$ \\ \hline
{\bf BBA} & $\eta^{(n)} _{j}$ & $-\tilde{\eta} ^{(n)}_{j}$ \\ \hline 
{\bf BAB} & $\eta^{(n)} _{-j}$ & $\eta^{(n)} _{-j}-\tilde{\eta} ^{(n)}_{-j}$ \\ \hline
{\bf ABB} & $-\eta^{(n)} _{j}$ & $-\eta^{(n)} _{j}+\tilde{\eta}^{(n)} _{j}$ \\ \hline
{\bf BBB} & $\eta^{(n)} _{j}$ & $\eta^{(n)} _{j}-\tilde{\eta}^{(n)} _j$ \\ \hline \hline
\end{tabular}
\end{center} 
\caption{ \label{Repsij} ${\cal R}$-images of the  $\theta ^n$-sector  ($n=2,4$) exceptional 3-cycles. 
 Note that the twist sector $n$ is preserved 
\cite{Blumenhagen:2002wn} under the orientifold action ${\cal R}$.}
\end{table}
 Similarly, for the  $\theta ^3$-sector exceptional cycles
 $\epsilon _j$ and $\tilde{\epsilon} _j$ with $j=1,4,5,6$, defined in (\ref{epsj}) and (\ref{epstilj}), the orientifold images are given 
  in Table \ref{Reps2} on all eight lattices. 
\begin{table}
 \begin{center}
\begin{tabular}{||c||c|c|c|c|c|c|c|c||} \hline \hline
 Lattice & ${\cal R} \epsilon _1$ &${\cal R} \tilde{\epsilon} _1$&${\cal R} \epsilon _4$&${\cal R} \tilde{\epsilon} _4$
& ${\cal R} \epsilon _5$ &${\cal R} \tilde{\epsilon} _5$ & ${\cal R} \epsilon _6$ &${\cal R} \tilde{\epsilon} _6$ \\ \hline \hline

{\bf AAA} & $-\epsilon _1-\tilde{\epsilon} _1 $& $\tilde{\epsilon} _1 $ &$-\epsilon _4-\tilde{\epsilon} _4 $& $\tilde{\epsilon} _4 $ 
&$-\epsilon _5-\tilde{\epsilon} _5 $& $\tilde{\epsilon} _5 $ &$-\epsilon _6-\tilde{\epsilon} _6 $& $\tilde{\epsilon} _6 $  \\ \hline

{\bf AAB} & $-\epsilon _1-\tilde{\epsilon} _1 $& $\tilde{\epsilon} _1 $ &$-\epsilon _4-\tilde{\epsilon} _4 $& $\tilde{\epsilon} _4 $ 
&$-\epsilon _6-\tilde{\epsilon} _6 $& $\tilde{\epsilon} _6 $ &$-\epsilon _5-\tilde{\epsilon} _5 $& $\tilde{\epsilon} _5 $ \\ \hline

{\bf ABA} & $-\epsilon _1$& $\epsilon _1 +\tilde{\epsilon} _1 $ &$-\epsilon _4 $& $\epsilon _4 +\tilde{\epsilon} _4 $ 
&$-\epsilon _5 $& $\epsilon _5+\tilde{\epsilon} _5 $ &$-\epsilon _6 $& $\epsilon _6+\tilde{\epsilon} _6 $ \\ \hline

{\bf BAA} & $\epsilon _1$& $-\epsilon _1 -\tilde{\epsilon} _1 $ &$\epsilon _4 $& $-\epsilon _4 -\tilde{\epsilon} _4 $ 
&$\epsilon _5 $& $-\epsilon _5-\tilde{\epsilon} _5 $ &$\epsilon _6 $& $-\epsilon _6-\tilde{\epsilon} _6 $ \\ \hline

{\bf ABB} & $-\epsilon _1$& $\epsilon _1 +\tilde{\epsilon} _1 $ &$-\epsilon _4 $& $\epsilon _4 +\tilde{\epsilon} _4 $ 
&$-\epsilon _6 $& $\epsilon _6+\tilde{\epsilon} _6 $ &$-\epsilon _5 $& $\epsilon _5+\tilde{\epsilon} _5 $ \\ \hline

{\bf BAB} & $\epsilon _1$& $-\epsilon _1 -\tilde{\epsilon} _1 $ &$\epsilon _4 $& $-\epsilon _4 -\tilde{\epsilon} _4 $ 
&$\epsilon _6 $& $-\epsilon _6-\tilde{\epsilon} _6 $ &$\epsilon _5 $& $-\epsilon _5-\tilde{\epsilon} _5 $ \\ \hline

{\bf BBA} & $-\tilde{\epsilon} _1 $& $-\epsilon _1 $ &$-\tilde{\epsilon} _4 $& $-\epsilon _4 $ 
&$-\tilde{\epsilon} _5 $& $-\epsilon _5 $ &$-\tilde{\epsilon} _6 $& $-\epsilon _6 $ \\ \hline 

{\bf BBB} & $-\tilde{\epsilon} _1 $& $-\epsilon _1 $ &$-\tilde{\epsilon} _4 $& $-\epsilon _4 $ 
&$-\tilde{\epsilon} _6 $& $-\epsilon _6 $ &$-\tilde{\epsilon} _5 $& $-\epsilon _5 $ \\ \hline \hline
\end{tabular}
\end{center} 
\caption{ \label{Reps2} ${\cal R}$-images of the  $\theta ^3$-sector exceptional 3-cycles.}
 \end{table}
Note that  on the {\bf ABA} lattice the action of ${\cal R}$ is just minus  the action on the {\bf BAA} lattice. 
Similarly for the {\bf ABB} and {\bf BAB} lattices.

The overall homology class for all exceptional branes and their orientifold images is required to vanish separately, since 
the O6-plane does not contribute. Thus
\beq
\sum _aN_a(\Pi _a^{\rm ex} +{\Pi _a^{\rm ex}}')=0 \label{tadex}
\eeq
We write the general exceptional brane (in the $\theta ^3$-sector) in the form
\beq
\Pi _a^{\rm ex}=\sum _{j=1,4,5,6}(\alpha ^a _j \epsilon _j+\tilde{\alpha}^a_j \tilde{\epsilon}_j)
\label{theta3ex}
\eeq
Using the results given in Table \ref{Reps2} the tadpole cancellation conditions (\ref{tadex}) are:
\bea
{\bf AAA}: &\quad& \sum _a N_a(\alpha ^a_j-2\tilde{\alpha}^a _j)=0  \label{RRexaaa}\\
{\bf AAB}: &\quad& \sum _a N_a(\alpha ^a_p-2\tilde{\alpha}^a _p)=0, \ 
\sum_aN_a\alpha ^a_5= \sum_aN_a(\tilde{\alpha}^a_5+ \tilde{\alpha}^a_6)=\sum_aN_a\alpha ^a_6\\
{\bf ABA}: &\quad& \sum _a N_a\tilde{\alpha}^a _j=0 \\
{\bf BAA}: &\quad& \sum _a N_a(2\alpha ^a_j-\tilde{\alpha}^a _j)=0 \\
{\bf ABB}: &\quad& \sum _a N_a\tilde{\alpha}^a _p=0, \ 
\sum_aN_a\tilde{\alpha} ^a_5= \sum_aN_a(\alpha^a_5- \alpha^a_6)=-\sum_aN_a\tilde{\alpha} ^a_6\\
{\bf BAB}: &\quad& \sum _a N_a(2\alpha ^a_p-\tilde{\alpha}^a _p)=0, \ 
\sum_aN_a\tilde{\alpha} ^a_5= \sum_aN_a(\alpha^a_5+ \alpha^a_6)=\sum_aN_a\tilde{\alpha} ^a_6\\
{\bf BBA}: &\quad& \sum _a N_a(\alpha ^a_j-\tilde{\alpha}^a _j)=0 \\
{\bf BBB}: &\quad& \sum _a N_a(\alpha ^a_p-\tilde{\alpha}^a _p)=0=
 \sum_aN_a(\alpha^a_5-\tilde{\alpha}^a_6)=\sum_aN_a(\alpha^a_6-\tilde{\alpha}^a_5)
\label{RRexbbb}
\eea
where $j=1,4,5,6$ and $p=1,4$. In each case there are 4 constraints, there being 4 independent $\mathcal{R}$-invariant combinations 
of the $\epsilon _j$ and $\tilde{\epsilon} _j$.
\section{Supersymmetric bulk 3-cycles} \label{susy}
 The twist (\ref{z61vk}) ensures that the closed-string sector is supersymmetric. In order to avoid supersymmetry breaking in the open-string sector, 
the D6-branes must wrap special Lagrangian cycles. Then the  stack $a$
 with wrapping numbers $(n^a_k,m^a_k) \ (k=1,2,3)$ is supersymmetric if 
\beq
\sum _{k=1}^3 \phi^a _k= 0 \bmod 2\pi \label{sumphik}
\eeq
where $\phi^a _k$ is the angle that the 1-cycle in $T^2_k$ makes with the Re $z_k$ axis. Thus
\beq
\phi^a _k = {\rm arg} (n^a_k e_{2k-1} +m^a_ke_{2k})
\eeq
where $e_{2k-1}$ and $e_{2k}$ are complex numbers defining the basis 1-cycles in $T^2_k$. 
Then, defining
\beq
Z^a \equiv \prod _{k=1}^3e_{2k-1}(n^a_k+m^a_kU_k) \equiv X^a+iY^a
\eeq
where
\beq
U_k \equiv \frac{e_{2k}}{e_{2k-1}}
\eeq
is the complex structure on $T^2_k$, 
the condition (\ref{sumphik}) that $a$ is  supersymmetric may be written as
\beq
X^a>0, \  Y^a=0 \label{XaYa}
\eeq
(A stack with $Y^a=0$ but $X^a<0$, so that $\sum _k \phi^a _k= \pi \bmod 2\pi$, corresponds to a (supersymmetric) stack of anti-D-branes.)   
In our case $T^2_{1,2}$ are $SU(3)$ lattices and the values of $U_k$ are given in (\ref{tau12}) and (\ref{tau3}). Thus
\beq
 Z^a= e_1e_3e_5[A^a_1-A^a_3+U_3(A^a_4-A^a_6)+e^{i\pi /3}(A^a_3+A^a_6U_3)]
 \eeq
 where $A^a_p \ (p=1,3,4,6)$ are the coefficients of the bulk 3-cycle defined in (\ref{A1})...(\ref{A6}).
If $T^2_k$ is of {\bf A}-type, $e_{2k-1}=R_{2k-1}>0$ is real and positive; in fact this is  also the case for $e_5$ 
when $T^2_3$ is of {\bf B}-type. However, when $T^2_k \ (k=1,2)$ is of {\bf B}-type, $e_{2k-1}=R_{2k-1}e^{-i\pi/6}$. It is 
then straightforward 
to evaluate  $X^a$ and $Y^a$ for the different lattices.
The results are given in Table \ref{susy3cycle}.
\begin{table}
 \begin{center}
\begin{tabular}{||c||c|c||} \hline \hline
 Lattice &  $X^a$ & $Y^a$ \\ \hline \hline
{\bf AAA}& $2A^a_1-A^a_3-A^a_6\sqrt{3}{\rm Im} \ U_3$ &$\sqrt{3}A^a_3+(2A^a_4-A^a_6){\rm Im} \ U_3 $ \\ \hline
{\bf AAB}&$ 2A^a_1-A^a_3+A^a_4-\frac{1}{2}A^a_6-A^a_6\sqrt{3}{\rm Im} \ U_3 $  & $\sqrt{3}(A^a_3+\frac{1}{2}A^a_6)+(2A^a_4-A^a_6){\rm Im} \ U_3 $ \\ \hline
{\bf ABA} and {\bf BAA} &$\sqrt{3}A^a_1+(A^a_4-2A^a_6){\rm Im} \ U_3 $& $2A^a_3-A^a_1+A^a_4\sqrt{3}{\rm Im} \ U_3$ \\ \hline
{\bf ABB} and {\bf BAB} &$\sqrt{3}(A^a_1+\frac{1}{2}A^a_4)+(A^a_4-2A^a_6){\rm Im} \ U_3 $& $2A^a_3-A^a_1+A^a_6-\frac{1}{2}A^a_4+A^a_4\sqrt{3}{\rm Im} \ U_3$ \\ \hline
{\bf BBA} &$ A^a_1+A^a_3+(A^a_4-A^a_6)\sqrt{3}{\rm Im} \ U_3$& $\sqrt{3}(A^a_3-A^a_1)+(A^a_4+A^a_6){\rm Im} \ U_3$ \\ \hline
{\bf BBB} &$(A^a_3+A^a_1+\frac{1}{2}A^a_6+\frac{1}{2}A^a_4)+$& $\sqrt{3}(A^a_3-A^a_1+\frac{1}{2}A^a_6-\frac{1}{2}A^a_4)$ \\ 
&$+(A_4-A_6)\sqrt{3}{\rm Im} \ U_3 $ & $+(A_4+A_6){\rm Im} \ U_3$ \\ \hline \hline
\end{tabular}
\end{center} 
\caption{ \label{susy3cycle} The functions $X^a$ and $Y^a$. (An overall  positive factor of $R_1R_3R_5$ is 
omitted.)  A  stack $a$ of 
D6-branes is supersymmetric if  $X^a>0$ and $Y^a=0$.}
 \end{table}

It follows that all of the ${\cal R}$-invariant bulk 3-cycles  given in Table \ref{O6Z61planes} satisfy $Y^a=0$, 
and (with a suitable choice of overall sign) are therefore supersymmetric. However,  
 unlike in the case of $\mathbb{Z} _6$, there are supersymmetric 3-cycles that are not ${\cal R}$-invariant. 
In particular, there are supersymmetric 3-cycles that do not wrap the O6-planes.
Thus, in general, $\Pi _a \circ \Pi _{\rm O6} \neq 0$ for the \Z$_6'$ orientifold. 
This has important implications which we will discuss shortly. 
 Further, the tadpole cancellation conditions (\ref{RRbulkaaa})  - (\ref{RRbulkbbb}) 
 allow an infinite number of solutions in this case. This is because the  two independent conditions for each lattice
  involve  independent linear combinations of 
the bulk coefficients $A^a_p$, whereas (for a fixed value of ${\rm Im} \  U_3$)
 the positivity condition required by supersymmetry, $X^a>0$ in (\ref{XaYa}), constrains just a single
 combination of these 
two; the orthogonal combination is unconstrained, and the condition $Y^a=0$ involves an additional, independent combination of the bulk coefficients.
 The combination of the tadpole cancellation conditions corresponding to this unconstrained combination of the bulk coefficients may therefore 
be satisfied by cancelling positive against 
negative contributions from different stacks, thereby allowing an infinite number of solutions.

\section{Fractional branes}
The bulk 3-cycles ($\Pi _a^{\rm bulk}$) or exceptional cycles ($\Pi _a^{\rm ex}$)
 discussed in \S\S \ref{z6'},\ref{z6'orientifold} cannot be used alone for realistic
 phenomenology, since their intersection numbers are always even, as is apparent from (\ref{pia0pib}),(\ref{etaj0k}) and (\ref{eps0jk}).
 To get odd intersection numbers, which,
as noted in \S \ref{intro}, is required if there is to be no unwanted additional vector-like matter,
 it is necessary to use fractional branes 
of the form
\beq
a= \frac{1}{2} \Pi _a^{\rm bulk}+ \frac{1}{2} \Pi _a^{\rm ex} \label{pifrac}
\eeq
$\Pi _a^{\rm bulk}=\sum _p A_p \rho _p$ is associated with wrapping numbers $(n^a_1,m^a_1)(n^a_2,m^a_2)(n^a_3,m^a_3)$,
 as shown in (\ref{genbulk}). The 
exceptional branes in the $\theta ^3$ sector are associated with the fixed points $f_{i,j}, \ (i,j=1,4,5,6)$ in $T^2_1 \otimes T^2_3$,
 as shown in 
(\ref{epsj}) and (\ref{epstilj}). If $\Pi _a^{\rm bulk}$ is a supersymmetric bulk 3-cycle, then the fractional brane $a$,
 defined in (\ref{pifrac}), preserves supersymmetry 
provided that the exceptional part arises only from fixed points traversed by the bulk 3-cycle. 
In what follows we shall only consider fractional branes whose 
exceptional part $\Pi _a^{\rm ex}$ arises from the $\theta ^3$-sector
 exceptional cycles $\epsilon _j$ and $\tilde{\epsilon}_j$ of the form (\ref{theta3ex}). 
 It appears to us that the $\theta ^2$-sector exceptional cycles $\eta _j$ and $\tilde{\eta} _j$, defined in 
 (\ref{etaj}), do not offer a rich enough structure to allow us 
 to satisfy all of the constraints that we shall impose.

As an illustration of the stacks  that we shall be considering, 
suppose that $(n^a_1,m^a_1)(n^a_3,m^a_3)=(1,1)(0,1) \bmod 2$. If the cycle on $T^2_1$ passes through 
 any two lattice points, for example the origin and $n^a_1e_1+m^a_1e_2$, then it wraps the $\mathbb{Z}_2$ fixed points 1 and 6,
defined in (\ref{1456}). If 
 the cycle on $T^2_1$ is offset from this situation  
  by one half of one of the  basis lattice vectors ($e_{1,2}$), for example  
   if it passes through $\frac{1}{2}e_1$ and $(n^a_1+\frac{1}{2})e_1+m^a_1e_2$, 
  then it wraps the $\mathbb{Z}_2$ fixed points 4 and 5. 
However, if the cycle is offset by  one half of both basis lattice vectors, for example if it passes through 
$\frac{1}{2}(e_1+e_2)$ and $(n^a_1+\frac{1}{2})e_1+(m^a_1+\frac{1}{2})e_2$, then it again wraps  the fixed points 1 and 6. 
Similarly, depending on the offset, the cycle on $T^2_3$ may wrap 1 and 5, or 4 and 6. 
Writing the offset in the form 
$\sum _{i=1,2} \sigma _i e _i\otimes \sum_{j=5,6}\sigma _je _j$ with $\sigma _{i,j} = 0, \frac{1}{2}$,
 the  exceptional cycles involved are those 
associated with the fixed points given in Table \ref{odododev}.
\begin{table}
 \begin{center}
\begin{tabular}{||c||c|c||} \hline \hline
$(\sigma _1, \sigma _2)$ & $(0,0)$ or $(\frac{1}{2}, \frac{1}{2})$ & $(\frac{1}{2},0)$ or $(0, \frac{1}{2})$ \\ \hline \hline
$(\sigma _5, \sigma _6)$  & & \\ \hline
$(0,0)$ or $(0, \frac{1}{2})$ & $f_{11},f_{15},f_{61},f_{65}$ & $f_{41},f_{45},f_{51},f_{55}$  \\ \hline
$(\frac{1}{2},0)$ or $(\frac{1}{2}, \frac{1}{2})$ & $f_{14},f_{16},f_{64},f_{66}$ & $f_{44},f_{46},f_{54},f_{56}$ \\ \hline \hline
\end{tabular}
\end{center} 
\caption{ \label{odododev} The fixed points used to generate $\theta ^3$-sector exceptional cycles
 in the case $(n^a_1,m^a_1)(n^a_3,m^a_3)=(1,1)(0,1) \bmod 2$.}
 \end{table}
Obviously a similar analysis applies for other choices of $(n^a_1,m^a_1)(n^a_3,m^a_3) \bmod 2$. 

In general, supersymmetry requires that  the exceptional part $\Pi _a^{\rm ex}$ of $a$ derives from  four fixed points,  
$f_{i^a_1j^a_1},f_{i^a_1j^a_2},f_{i^a_2j^a_1},f_{i^a_2j^a_2}$, with  $(i^a_1,i^a_2)$ the fixed points in $T^2_1$, 
and $(j^a_1,j^a_2)$ those in $T^2_3$. Depending on the offset, there are two choices for $(i^a_1,i^a_2)$ in $T^2_1$, and similarly 
two choices for $(j^a_1,j^a_2)$ in $T^2_3$. Consequently there are a total of four sets of four fixed points, any one of 
which may be used to determine $\Pi _a^{\rm ex}$ while preserving supersymmetry. 
 The choice of Wilson lines affects the relative signs with which the contributions from the four fixed points are combined 
 to determine $\Pi _a ^{\rm ex}$. The rule is that
 \bea
 (i^a_1,i^a_2)(j^a_1,j^a_2)& \rightarrow & f_{i^a_1j^a_1},f_{i^a_1j^a_2},f_{i^a_2j^a_1},f_{i^a_2j^a_2} \\
& \rightarrow &(-1)^{\tau^a _0} \left[ f_{i^a_1j^a_1}+(-1)^{\tau^a _2}f_{i^a_1j^a_2}+(-1)^{\tau^a _1}f_{i^a_2j^a_1}
 +(-1)^{\tau^a _1+\tau^a _2}f_{i^a_2j^a_2} \right]
 \eea
where $\tau^a _{0,1,2}=0,1$ with $\tau^a _1 =1$ corresponding to a Wilson line in $T^2_1$ and likewise for $\tau^a _2$ in $T^2_3$. 
The fixed point $f_{i,j} \ (i,j=1,4,5,6)$ with the 1-cycle $n^a_2\pi _3 +m^a_2\pi _4$ 
 is then associated with (orbifold-invariant) exceptional cycle as shown in Table \ref{FPex}. 
 Thus in the example above with $(n^a_1,m^a_1)(n^a_3,m^a_3)=(1,1)(0,1) \bmod 2$, if we take no offset in both planes, 
then $(i^a_1,i^a_2)=(16)$ and $(j^a_1,j^a_2)=(15)$. With this choice
 \beq
 \Pi _{a \ (16)(15)} ^{\rm ex}(n^a_2,m^b_2)=(-1)^{\tau^a _0 +\tau^a _1}\{ n^a_2 [\epsilon _1+(-1)^{\tau^a _2}\epsilon _5 ] -
m^a_2 [\tilde{\epsilon }_1+(-1)^{\tau^a _2}\tilde{\epsilon} _5 ] \} \label{pi1615}
\eeq

We noted earlier that given any two stacks $a$ and $b$ of D6-branes, then there is  chiral matter whose
 spectrum is  determined by the intersection 
numbers of the 3-cycles. There are $a \circ b$ chiral matter multiplets in the bifundamental representation 
 $({\bf N} _a, \overline{\bf N}_b)$ of $U(N_a) \otimes U(N_b)$, and $a \circ b'$ matter multiplets in the  representation 
 $({\bf N} _a, {\bf N}_b)$, where  $b' \equiv \mathcal{R}b$ is the orientifold image of $b$. In general, there is also  
 chiral matter in the symmetric ${\bf S}_a$ and antisymmetric representations ${\bf A}_a$ of the gauge group $U(N_a)$, and 
 likewise for $U(N_b)$, where the dimensions of these representations are
 \bea
\left [{\bf S} _a\right] &\equiv& ({\bf N} _a \times {\bf N} _a)_{\rm symm} = \frac{1}{2}N_a(N_a+1) \\
\left[{\bf A} _a\ \right] &\equiv& ({\bf N} _a \times {\bf N} _a)_{\rm antisymm} = \frac{1}{2}N_a(N_a-1)
\eea 
The number of multiplets in the ${\bf S}_a$ representation is $\frac{1}{2}(a \circ a' -a \circ \Pi _{\rm O6})$, and 
the number of multiplets in the ${\bf A}_a$ representation is $\frac{1}{2}(a \circ a' +a \circ \Pi _{\rm O6})$, where 
$\Pi _{\rm O6}$ is the total O6-brane homology class given  in Table \ref{pio6Z61}. If $a \circ \Pi _{\rm O6}\neq 0$, 
then copies of one or both representations are inevitably present. This explains the importance of our observation at the end of \S \ref{susy}
 that this quantity is generally non-zero on the \Z$_6'$ orientifold. Of course, for a specific stack, it may happen that  $a \circ \Pi _{\rm O6} =0$, 
 in which case  the absence of one of the representations ensures that neither representation is present.
 When $N_a=3$, as required to get the 
$SU(3)_c$ gauge group of QCD, there will in general be chiral matter in the ${\bf S}_a={\bf 6}$ and ${\bf A}_a =\overline{\bf 3}$
 of the  $SU(3)$. Similarly, when $N_b=2$, as required to get the 
 $SU(2)_L$ part of the electroweak gauge group, there will in general be chiral matter in the ${\bf S}_b={\bf 3}$
 and ${\bf A}_a =\overline{\bf 1}$ of  $SU(2)$. For both groups the appearance of the symmetric representation is excluded 
 phenomenologically. Quark singlet states $q^c_L$, in the $\overline{\bf 3}$ representation of $SU(3)_c$, 
do occur of course, and similarly for lepton singlet states $\ell ^c_L$ in the ${\bf 1}$ (or $\overline{\bf 1}$)
 representation of $SU(2)_L$. Thus, 
we must exclude the appearance of the representations ${\bf S}_a$ and ${\bf S}_b$, but not necessarily of ${\bf A}_a$ or ${\bf A}_b$.
Consequently, we impose the constraints
\bea
a \circ a'&=& a \circ \Pi _{\rm O6} \label{aa'o6}\\
b \circ b'&=& b \circ \Pi _{\rm O6} \label{bb'o6}
\eea
With these constraints the  number of multiplets in the antisymmetric representation ${\bf A}_a$ is then $a \circ \Pi _{\rm O6}$, 
so we require too that
\beq
 |a \circ \Pi _{\rm O6}|\leq 3  \label{aopio6}
 \eeq
  since otherwise there would again be non-minimal vector-like quark singlet matter. 
Similarly we require that
\beq
|b \circ \Pi _{\rm O6}|\leq 3
 \label{bopio6}
\eeq
to avoid unwanted vector-like lepton singets.

For fractional branes of the form (\ref{pifrac}),
\beq
a \circ a' = \frac{1}{4} \Pi^{\rm bulk} _a \circ \Pi^{\rm bulk \ '} _a+ \frac{1}{4} \Pi _a ^{\rm ex}  \circ \Pi^{\rm ex \ '} _a
\eeq 
and
\beq
a \circ \Pi _{\rm O6} =\frac{1}{2}\Pi^{\rm bulk} _a \circ \Pi _{\rm O6}
\eeq
since there is no intersection between the O6-planes and the exceptional branes. 
For given wrapping numbers  $(n^a_1,m^a_1)(n^a_2,m^a_2)(n^a_3,m^a_3)$, the form (\ref{genbulk}) of the bulk part $\Pi^{\rm bulk} _a$ 
of $a$  is determined by the formulae (\ref{A1})...(\ref{A6}) for the bulk coefficients $A^a_p  \ (p=1,3,4,6)$. For any given lattice,
it is then straightforward to calculate $\Pi^{\rm bulk \ '} _a$ using Table \ref{Rrhoz61} and hence, using (\ref{pia0pib}), to calculate 
\beq
f(A^a_p) \equiv  \Pi^{\rm bulk} _a \circ {\Pi^{\rm bulk \ } _a}'=F(A^a_p,{A^a_p}')
\eeq
where $A^a_p,{A^a_p}'$ are respectively the bulk coefficients for $\Pi^{\rm bulk} _a, \Pi^{\rm bulk \ '} _a$.
Similarly, using Table \ref{pio6Z61}, it is straightforward to calculate
\beq
g(A^a_p) \equiv \Pi^{\rm bulk} _a \circ \Pi _{\rm O6}=F(A^a_p,S_p) \label{gAp}
\eeq
where $S_p$ are the bulk coefficients of the O6-planes.
The calculation of 
\beq
\phi (n^a_2,m^a_2) \equiv  \Pi _a ^{\rm ex}  \circ {\Pi^{\rm ex \ } _a}' \label{fi}
\eeq
using (\ref{eps0jk}) is equally straightforward, 
but also depends upon which one of the four sets of four fixed points $(i^a_1,i^a_2)(j^a_1,j^a_2)$ is chosen. 
The constraint (\ref{aa'o6}) may then be written as
\beq 
f(A^a_p)+\phi(n_2^a,m_2^a)=2g(A^a_p) \label{fphig}
\eeq
and (\ref{aopio6}) as
\beq
|g(A^a_p) | \leq 6 \label{gAap6}
\eeq
Similarly, for the stack $b$ we require that
\beq 
f(A^b_p)+\phi(n_2^b,m_2^b)=2g(A^b_p) \label{fphigb}
\eeq
and 
\beq
|g(A^b_p) | \leq 6 \label{gAbp6}
\eeq

Each pair of wrapping numbers $(n^a_k,m^a_k)  \ (k=1,2,3)$ may take one of the three values \linebreak $(1,1),(1,0),(0,1) \bmod2$, 
since $n^a_k$ and $m^a_k$ are coprime. Thus, in general there are 27 possibilities for $(n^a_1,m^a_1)(n^a_2,m^a_2)(n^a_3,m^a_3)\bmod 2$. 
However, from (\ref{theta12}),(\ref{theta34}) and (\ref{thetaZ6}), it follows that 
the \Z$_6'$ point group generator $\theta$ acts on the wrapping numbers as
\beq
\theta (n^a_1,m^a_1)(n^a_2,m^a_2)(n^a_3,m^a_3) = (-m^a_1, n^a_1+m^a_1)(-n^a_2-m^a_2,n^a_2)(-n^a_3,-m^a_3)
\eeq
This reduces the 27 possibilities  to 9, since any choice is related to two others by the action of $\theta$. Since 
$(n^a_3,m^a_3)$ is invariant $\bmod 2$ under the action of $\theta$, we may choose a ``gauge'' in which the representative 
element in each class has  $(n^a_1,m^a_1)=(n^a_3,m^a_3) \bmod 2$.  
 The 9 classes may be reduced further by including the action 
   $\mathcal{R}$ of the world sheet parity operator, but the action depends upon the lattice choice.
An important difference  arises between the four lattices in which $T^2_3$ is {\bf A} type, and the four in which it is of {\bf B} type.
In the former case, it follows from (\ref{RA31}) that all 3 choices for $(n^a_3,m^a_3)=(1,1),(1,0),(0,1) \bmod 2$ are  $\mathcal{R}$ invariant. This means that the 
3 classes defined above are not mixed by the action of $\mathcal{R}$. The 3 elements within  each class split into $1+2$, where 1 
is $\mathcal{R}$ invariant and the 2 are related under the action of $\mathcal{R}$. Thus, on the 4 lattices with $T^2_3$ of 
{\bf A} type, the 3 classes of 3 reduce to 3 classes of 2, leaving 6 in total. In the latter case, when $T^2_3$  is of 
{\bf B} type, it follows from (\ref{RB31}) that $(n^a_3,m^a_3)=(1,0) \bmod 2$ is \R  \ invariant, and that 
$(1,1) \leftrightarrow (0,1)$ under the action of \R. 
This means that of the 9 classes there is just one that is \R \ invariant, and the remaining 8 split into 4 classes of 2 in 
which each pair is related by the action of \R. Thus the 9 classes reduce to 5. The results are summarised in Table \ref{nmlattices}.
\begin{table}
 \begin{center}
\begin{tabular}{||c||c|c|c|c||} \hline \hline
Lattice & $(n^a_1,m^a_1)(n^a_2,m^a_2)(n^a_3,m^a_3) \bmod 2$ &$f(A_i)$ &$\phi(n^a_2,m^a_2)$& $2g(A_i)$  \\ \hline \hline
{\bf AAA} & $(1,1)(1,1)(1,1) \leftrightarrow (1,1)(1,0)(1,1)$&4&4&0 \\ \cline{2-5}
&$(1,1)(0,1)(1,1)$&0&0&0 \\ \cline{2-5}
&$(1,0)(1,1)(1,0) \leftrightarrow (1,0)(0,1)(1,0)$  &0&4&4\\ \cline{2-5}
&$(1,0)(1,0)(1,0)$&0&0&0  \\ \cline{2-5}
&$(0,1)(1,0)(0,1) \leftrightarrow (0,1)(0,1)(0,1)$ &0&4&4\\ \cline{2-5}
&$(0,1)(1,1)(0,1)$ &0&0&0\\ \hline
{\bf BAA} and {\bf ABA} & $(1,1)(1,1)(1,1) \leftrightarrow (1,1)(0,1)(1,1)$&4&4&0 \\ \cline{2-5}
&$(1,1)(1,0)(1,1)$ &0&0&0\\ \cline{2-5}
&$(1,0)(1,0)(1,0) \leftrightarrow (1,0)(0,1)(1,0)$ &0&4&4\\ \cline{2-5}
&$(1,0)(1,1)(1,0)$ &0&0&0\\ \cline{2-5} 
&$(0,1)(1,1)(0,1) \leftrightarrow (0,1)(1,0)(0,1)$&0&4&4 \\ \cline{2-5}
&$(0,1)(0,1)(0,1)$ &0&0&0\\ \hline
{\bf BBA} & $(1,1)(1,0)(1,1) \leftrightarrow (1,1)(0,1)(1,1)$ &4&4&0\\ \cline{2-5}
&$(1,1)(1,1)(1,1)$ &0&0&0\\ \cline{2-5}
&$(1,0)(1,1)(1,0) \leftrightarrow (1,0)(1,0)(1,0)$ &0&4&4\\ \cline{2-5}
&$(1,0)(0,1)(1,0)$ &0&0&0\\ \cline{2-5} 
&$(0,1)(1,1)(0,1) \leftrightarrow (0,1)(0,1)(0,1)$ &0&4&4\\ \cline{2-5}
&$(0,1)(1,0)(0,1)$ &0&0&0\\ \hline \hline
{\bf AAB} &$(1,1)(1,1)(1,1) \leftrightarrow (0,1)(0,1)(0,1)$ &2&6&0\\ \cline{2-5}
&$(1,1)(1,0)(1,1) \leftrightarrow (0,1)(1,0)(0,1)$ &2&2&0\\ \cline{2-5}
&$(1,1)(0,1)(1,1) \leftrightarrow (0,1)(1,1)(0,1)$ &4&0&0\\ \cline{2-5}
&$(1,0)(1,1)(1,0) \leftrightarrow (1,0)(0,1)(1,0)$ &0&4&0\\ \cline{2-5}
&$(1,0)(1,0)(1,0)$ &0&0&0\\ \hline 
{\bf BAB} (and {\bf ABB}) &$(1,1)(1,1)(1,1) \leftrightarrow (0,1)(1,1)(0,1)$ &6&2(6)&0\\ \cline{2-5}
&$(1,1)(1,0)(1,1) \leftrightarrow (0,1)(0,1)(0,1)$&4&0&0 \\ \cline{2-5}
&$(1,1)(0,1)(1,1) \leftrightarrow (0,1)(1,0)(0,1)$ &6&6(2)&0\\ \cline{2-5}
&$(1,0)(1,0)(1,0) \leftrightarrow (1,0)(0,1)(1,0)$ &0&4&0\\ \cline{2-5}
&$(1,0)(1,1)(1,0)$ &0&0&0\\ \hline 
{\bf BBB} &$(1,1)(1,1)(1,1) \leftrightarrow (0,1)(1,0)(0,1)$ &4&0&0\\ \cline{2-5}
&$(1,1)(1,0)(1,1) \leftrightarrow (0,1)(1,1)(0,1)$ &2&2&0\\ \cline{2-5}
&$(1,1)(0,1)(1,1) \leftrightarrow (0,1)(0,1)(0,1)$ &2&6&0\\ \cline{2-5}
&$(1,0)(1,1)(1,0) \leftrightarrow (1,0)(1,0)(1,0)$ &0&4&0\\ \cline{2-5}
&$(1,0)(0,1)(1,0)$&0&0&0 \\ \hline \hline 
\end{tabular}
\end{center} 
\caption{ \label{nmlattices} Inequivalent wrapping number classes $(n^a_k,m^a_k)\bmod 2$ for the various lattices, and
the values of the functions $f(A_p)$, $\phi(n^a_2,m^a_2)$ and $2g(A_p) \bmod 8$.} 
\end{table}
The functions $f(A^a_p)$ and $g(A^a_p)$  are summarised in Table \ref{latfnfg}
\begin{table}
\begin{center}
\begin{tabular}{||c||c|c||} \hline \hline
 Lattice & $f(A^a_p)$ & $g(A^a_p)$  \\ \hline \hline
{\bf AAA} &$4(2A^a_1A^a_4-2A^a_1A^a_6-A^a_3A^a_6)$ &$2(2A^a_4-3A^a_3-A^a_6)$   \\ \hline
{\bf AAB} &$4(2A^a_1A^a_4-2A^a_1A^a_6-A^a_3A^a_6) +2(2{A^a_4}^2-2A^a_4A^a_6-{A^a_6}^2)$ &$4(3A^a_3+A^a_4+A^a_6)$   \\  \hline
{\bf ABA} &$4(A^a_1A^a_4+2A^a_1A^a_6-2A^a_3A^a_6)$ &$6(2A^a_3+A^a_4-A^a_1)$   \\ \hline
{\bf BAA} &$4(A^a_1A^a_4+2A^a_1A^a_6-2A^a_3A^a_6)$ &$2(A^a_1-2A^a_3+A^a_4)$  \\ \hline
{\bf ABB} &$4(A^a_1A^a_4+2A^a_1A^a_6-2A^a_3A^a_6) +2({A^a_4}^2+2A^a_4A^a_6-2{A^a_6}^2)$ &$12(A^a_1-2A^a_3+A^a_4-A^a_6)$    \\  \hline
{\bf BAB} &$4(A^a_1A^a_4+2A^a_1A^a_6-2A^a_3A^a_6)+2({A^a_4}^2+2A^a_4A^a_6-2{A^a_6}^2)$ &$4(-A^a_1+2A^a_3+A^a_4+A^a_6)$   \\  \hline
{\bf BBA} &$4(4A^a_1A^a_6-A^a_1A^a_4-A^a_3A^a_6)$ &$6(A^a_3-A^a_1+A^a_4+A^a_6)$   \\  \hline
{\bf BBB} &$-4(4A^a_1A^a_4-4A^a_1A^a_6+A^a_3A^a_6)+2(-{A^a_4}^2+4A^a_4A^a_6-{A^a_6}^2)$ &$12(A^a_1-A^a_3+A^a_4)$   \\  \hline \hline
\end{tabular}
\end{center} 
\caption{ \label{latfnfg} The functions $f(A^a_p)$ and  $g(A^a_p)$  for various lattices. } 
\addtolength{\evensidemargin}{15mm}
\addtolength{\oddsidemargin}{15mm}
\end{table}
 and the function $\phi(n^a_2,m^a_2)$ is given in 
Table \ref{latfnphi}.
\addtolength{\evensidemargin}{-2\evensidemargin}
\addtolength{\oddsidemargin}{-2\oddsidemargin}
\begin{table}
\begin{center}
\begin{tabular}{||c||c|c|c||} \hline \hline
Lattice & $(1,1)$ & $(1,0)$ & $(0,1)$  \\ \hline \hline
{\bf AAA} & $4n^a_2(n^a_2+2m^a_2)$ & $-4m^a_2(m^a_2+2n^a_2)$ & $4({m^a_2}^2-{n^a_2}^2)$  \\ \hline
{\bf AAB} & $2n^a_2(n^a_2+2m^a_2)$ & $-4m^a_2(m^a_2+2n^a_2)$ & $2({m^a_2}^2-{n^a_2}^2)$  \\ \hline
{\bf ABA} & $4m^a_2(m^a_2+2n^a_2)$&$-4({m^a_2}^2-{n^a_2}^2)$ & $-4n^a_2(n^a_2+2m^a_2)$  \\ \hline
{\bf BAA} & $-4m^a_2(m^a_2+2n^a_2)$ &  $4({m^a_2}^2-{n^a_2}^2)$ & $4n^a_2(n^a_2+2m^a_2)$  \\ \hline
{\bf ABB} & $2m^a_2(m^a_2+2n^a_2)$ &  $-4({m^a_2}^2-{n^a_2}^2)$ & $-2n^a_2(n^a_2+2m^a_2)$   \\ \hline
{\bf BAB} & $-2m^a_2(m^a_2+2n^a_2)$ &  $4({m^a_2}^2-{n^a_2}^2)$ & $2n^a_2(n^a_2+2m^a_2)$   \\ \hline
{\bf BBA}  &$-4({m^a_2}^2-{n^a_2}^2)$ & $-4n^a_2(n^a_2+2m^a_2)$ & $4m^a_2(m^a_2+2n^a_2)$   \\  \hline
{\bf BBB} & $-2({m^a_2}^2-{n^a_2}^2)$ & $-4n^a_2(n^a_2+2m^a_2)$ & $2m^a_2(m^a_2+2n^a_2)$   \\ \hline \hline
\end{tabular}
\end{center} 
\caption{ \label{latfnphi} The function $\phi(n^a_2,m^a_2)$ for various lattices. The three 
values for $\phi$ correspond to the three values of $(n^a_1,m^a_1)=(n^a_3,m^a_3)\bmod2$ when the 
singular points chosen  in both tori include the origin.} 
\addtolength{\evensidemargin}{15mm}
\addtolength{\oddsidemargin}{15mm}
\end{table}
Their values ($\bmod 8$)
 are given in Table \ref{nmlattices}. By inspection we can see that the requirement (\ref{fphig}) that there are no 
 symmetric representations places no further restrictions ($\bmod \ 8$) on the allowed wrapping numbers when $T^2_3$ is of {\bf A} type. 
 However, when $T^2_3$ is of {\bf B} type, (\ref{fphig}) limits the possible wrapping numbers to just 2 of the 5 classes 
 allowed hitherto.

As noted earlier, the function 
$\phi(n^a_2,m^a_2)$ defined in (\ref{fi}) depends upon which pairs of fixed points 
 are used in $T^2_1$ and $T^2_3$, 
although we have not displayed this dependence. 
Table \ref{latfnphi} gives the results 
when the  (non-offset) pair (16) is chosen in both tori in the $(n^a_1,m^a_1)=(n^a_3,m^a_3)=(1,1) \bmod 2$ case, when (14) is chosen in the $(1,0)$ 
case, and when (15) is chosen on the $(0,1)$ case. The modifications that arise when the other pair of fixed points, 
the offset pair,
 is chosen in one or both tori may be summarised as follows. For the four lattices {\bf AAA, ABA, BAA, BBA} in which $T^2_3$ is 
 of {\bf A} type, there is no modification when the offset pair is chosen in $T^2_3$, but an additional 
 multiplicative  factor of $2(-1)^{\tau ^a_1}-1$ is to be inserted when the offset pair is used in $T^2_1$. 
For the four lattices {\bf AAB, ABB, BAB, BBB} in which $T^2_3$ is 
 of {\bf B} type, the four different choices of pairs of fixed points can lead to different results. When the offset pair 
 is used in  $T^2_3$ but not in $T^2_1$, a  multiplicative factor of $(-1)^{\tau ^a_2}$ is to be inserted, 
 but {\em only} in the case $(n^a_1,m^a_1)=(n^a_3,m^a_3) =(1,0) \bmod 2$.  
When the offset pair 
 is used in  $T^2_1$ but not in $T^2_3$, a  multiplicative factor of $2(-1)^{\tau ^a_1}-1$  is to be inserted in all cases.  
When the offset pair 
 is used in   both $T^2_1$ and $T^2_3$, both factors are to be inserted. Since $2(-1)^{\tau ^a_1}-1$ and $(-1)^{\tau ^a_2}$
 are both odd, their inclusion does not affect our conclusion that when $T^2_3$ is of {\bf A} type
  the allowed wrapping numbers are not further restricted ($\bmod 8$) by 
 equation (\ref{fphig}). In other words, for these lattices, the allowed classes of wrapping numbers are unaffected by 
 the inclusion of Wilson lines. 

However, when $T^2_3$ is of {\bf B} type, the allowed classes of wrapping numbers 
 could be affected by Wilson lines. A difference could arise whenever $\phi(n^a_2,m^a_2) = \pm 2 \bmod 8$. Insertion of a 
 factor  $2(-1)^{\tau ^a_1}-1$ has no effect since $2=-6\bmod 8$, but insertion of a factor  $(-1)^{\tau ^a_2}$ 
 clearly {\em does} have an effect. However,  $\phi(n^a_2,m^a_2) = \pm 2 \bmod 8$ only occurs when $(n^a_1,m^a_1)=
 (n^a_3,m^a_3)=(1,1) \bmod 2$, and the factor of $(-1)^{\tau _2^a}$ does not then occur, as noted above. Thus, for the four lattices 
 {\bf AAB, ABB, BAB, BBB} in which $T^2_3$ is of {\bf B} type also, the (two) allowed classes of wrapping numbers are unaffected 
 by the inclusion of Wilson lines.
 In this case, there are 3 possibilities for $a$ and $b$. 
 In two of them, $a$ and $b$ are in the same class; in the third, they 
are in different classes. In the last of these possibilities, it is convenient to take the stack $a$ to be that in which  $(n^a_1,m^a_1)=(n^a_3,m^a_3)=(1,0) \bmod 2$  and 
 $(n^b_1,m^b_1)=(n^b_3,m^b_3)=(1,1) \bmod 2$.  Consequently we may no longer assume that $N_a=3$ and $N_b=2$, as we have done hitherto. 

In this paper 
we see whether we can construct realistic models, that is models in which the intersection numbers
$(|a \circ b|, |a \circ b'|)=(2,1)$ or $(1,2)$, using the 3 possible combinations of 
 wrapping numbers for each of the 4 lattices in which $T^2_3$ is of {\bf B} type. 
 As noted in the introduction, if $a \circ b$ and $a \circ b'$ have the same sign, then $N_a=3$ and $N_b=2$. If not, $N_a=2$ and $N_b=3$. 
The contributions from the bulk parts to $a \circ b$ and $a \circ b'$ are determined by the functions $f_{AB}$ and $f_{AB'}$ respectively.
 The former is independent of the lattice and is given by 
\bea
f_{AB} \equiv \Pi_a^{\rm bulk}\circ \Pi_b^{\rm bulk}= F(A^a_p,A^b_p)   \label{fAB}
\eea
where $F(A^a_p,A^b_p)$ is defined in (\ref{pia0pib}). 
The latter,
\beq
f_{AB'}\equiv \Pi_a^{\rm bulk}\circ {\Pi_b^{\rm bulk}}'= F(A^a_p,{A^b_p}')   \label{fABprime}
\eeq
 however, {\em is} lattice-dependent. 
 Table \ref{fAB1} gives the function $f_{AB'}$ for each lattice.
 \begin{table}
 \begin{center}
\begin{tabular}{||c||c||} \hline \hline
Lattice & $f_{AB'}(A^a_p,A^b_p)$ \\ \hline \hline
{\bf AAA} &$4(A^a_1A^b_4+A^a_4A^b_1-2(A^a_1A^b_6+A^a_6A^b_1)-2(A^a_3A^b_4+A^a_4A^b_3)-2(A^a_3A^b_6+A^a_6A^b_3)$ \\ \hline
{\bf ABA} and {\bf BAA} & $2(A^a_1A^b_4+A^a_4A^b_1)+2(A^a_1A^b_6+A^a_6A^b_1)+2(A^a_3A^b_4+A^a_4A^b_3)-4(A^a_3A^b_6+A^a_6A^b_3)$ \\ \hline
{\bf BBA}& $-2(A^a_1A^b_4+A^a_4A^b_1)+4(A^a_1A^b_6+A^a_6A^b_1)+4(A^a_3A^b_4+A^a_4A^b_3)-2(A^a_3A^b_6+A^a_6A^b_3)$ \\ \hline
{\bf AAB} & $4(A^a_1A^b_4+A^a_4A^b_1)+4A^a_4A^b_4-2(A^a_1A^b_6+A^a_6A^b_1)-2(A^a_3A^b_4+A^a_4A^b_3)-$ \\
&$-2(A^a_3A^b_6+A^a_6A^b_3)-2A^a_6A^b_6-2(A^a_4A^b_6+A^a_6A^b_4)$ \\ \hline
{\bf ABB} and {\bf AAA} & $2(A^a_1A^b_4+A^a_4A^b_1)+2A^a_4A^b_4+2(A^a_1A^b_6+A^a_6A^b_1)+2(A^a_3A^b_4+A^a_4A^b_3)-$\\
&$-4(A^a_3A^b_6+A^a_6A^b_3)-4A^a_6A^b_6+2(A^a_4A^b_6+A^a_6A^b_4)$ \\ \hline
{\bf BBB} &$-2(A^a_1A^b_4+A^a_4A^b_1)-2A^a_4A^b_4+4(A^a_1A^b_6+A^a_6A^b_1)+4(A^a_3A^b_4+A^a_4A^b_3)-$ \\
&$-2(A^a_3A^b_6+A^a_6A^b_3)-2A^a_6A^b_6+4(A^a_4A^b_6+A^a_6A^b_4)$ \\ \hline \hline
\end{tabular}
\end{center} 
\caption{ \label{fAB1} The function $f_{AB'}$, defined in (\ref{fABprime}), for the various lattices. } 
\end{table}
The contribution to $a \circ b$ and $a \circ b'$ from the exceptional parts depends upon the fixed points chosen for 
$\Pi ^{\rm ex}_a$ and $\Pi ^{\rm ex}_b$. Since there are four possible sets of fixed points for each exceptional part, 
there are, in principle 10 or 16 combinations to consider, depending upon whether $a$ and $b$ have the same or different 
sets of wrapping numbers ($\bmod \ 2$). 
The required intersection numbers are then
\bea
a \circ b&=& \frac{1}{4}f_{AB} + \frac{1}{4} \Pi^{\rm ex}_{a \ (i^a_1,i^a_2)(j^a_1,j^a_2)}(n^a_2,m^a_2) \circ 
\Pi^{\rm ex}_{b \ (i^b_1,i^b_2)(j^b_1,j^b_2)}(n^b_2,m^b_2) \label{aob}\\
a \circ b'&=& \frac{1}{4}f_{AB'} + \frac{1}{4} \Pi^{\rm ex}_{a \ (i^a_1,i^a_2)(j^a_1,j^a_2)}(n^a_2,m^a_2) \circ 
{\Pi^{\rm ex}_{b \ (i^b_1,i^b_2)(j^b_1,j^b_2)}}'(n^b_2,m^b_2) \label{aob'}
\eea
where we are using the notation of  equation (\ref{pi1615}). In what follows we abbreviate this notation by
\bea
\Pi^{\rm ex}_{a \ (i^a_1,i^a_2)(j^a_1,j^a_2)}(n^a_2,m^a_2)\rightarrow (i^a_1,i^a_2)(j^a_1,j^a_2)\\
{\Pi^{\rm ex}_{a \ (i^a_1,i^a_2)(j^a_1,j^a_2)}}'(n^a_2,m^a_2)\rightarrow (i^a_1,i^a_2)(j^a_1,j^a_2)'
\eea
Like the contribution $f_{AB}$ from the bulk branes to $a \circ b$, the contribution  from the exceptional branes is 
independent of the lattice, and depends only on the  associated fixed points. We have seen that on all four lattices in which $T^2_3$ is of 
{\bf B} type, the 
requirement that there is no matter in symmetric representations of the gauge group means that we must consider three cases: 
 (i) $(n^{a,b}_1, m^{a,b}_1)=(n^{a,b}_3, m^{a,b}_3)=(1,1) \bmod 2$, 
(ii)  $(n^{a,b}_1, m^{a,b}_1)=(n^{a,b}_3, m^{a,b}_3)=(1,0) \bmod 2$, and 
(iii)  $(n^{a}_1, m^{a}_1)=(n^{a}_3, m^{a}_3)=(1,0) \bmod 2$, $(n^{b}_1, m^{b}_1)=(n^{b}_3, m^{b}_3)=(1,1) \bmod 2$.
Thus exceptional branes associated with the same fixed points arise on all four lattices. 
 In the next section we 
give the (universal) results for the contributions of these exceptional branes to  $a \circ b$. 
(As is apparent from Table \ref{nmlattices}, the allowed values of $(n^{a,b}_2, m^{a,b}_2) \bmod 2$ {\em are} lattice-dependent, but
 we present the results for arbitrary values of $(n^{a,b}_2, m^{a,b}_2)$.)
 
The corresponding contributions from the exceptional branes to $a \circ b'$ are presented for each lattice in the 
appendices. 
In all cases we give the results for the case that $(i^a_1,i^a_2)$ and $(i^b_1,i^b_2)$ are the offset  pairs  of fixed points in $T^2_1$; 
the results for the cases when one or both pairs are not offset are obtained by setting $\tau ^a_1=0$ and/or $\tau ^b_1=0$.
\section{ Calculations of $(i^a_1,i^a_2)(j^a_1,j^a_2)\circ (i^b_1,i^b_2)(j^b_1,j^b_2)$}
\subsection{$(n^{a,b}_1, m^{a,b}_1)=(n^{a,b}_3, m^{a,b}_3)=(1,1) \bmod 2$} \label{11}
In this case $(i^a_1,i^a_2),(i^b_1,i^b_2)=(45)$ and $(j^a_1,j^a_2),(j^b_1,j^b_2)=(16)$ or $(45)$.
\bea
(45)(16) &\circ &  (45)(16) = (45)(45) \circ (45)(45) =\nonumber \\
&=& (-1)^{\tau ^a _0 +\tau ^b _0 +1}2[1+(-1)^{\tau ^a_2 +\tau ^b_2}][m^a_2n^b_2-n^a_2m^b_2 + \nonumber \\
&+&(-1)^{\tau ^a_1+1}(n^a_2n^b_2+m^a_2m^b_2+m^a_2n^b_2)+(-1)^{\tau ^b_1}(n^a_2n^b_2+m^a_2m^b_2+n^a_2m^b_2) + \nonumber \\
&+&(-1)^{\tau ^a _1 +\tau ^b_1}(m^a_2n^b_2-n^a_2m^b_2)] \\
(45)(16) &\circ &  (45)(45) =0 
\eea
\subsection{$(n^{a,b}_1,m^{a,b}_1)=(n^{a,b}_3,m^{a,b}_3)=(1,0)  \bmod 2$} \label{10}
In this case $(i^a_1,i^a_2),(i^b_1,i^b_2)=(56)$ and $(j^a_1,j^a_2),(j^b_1,j^b_2)=(14)$ or $(56)$.
\bea
(56)(14) &\circ &  (56)(14) =(56)(56) \circ (56)(56) =\nonumber \\
&=& (-1)^{\tau ^a _0 +\tau ^b _0 +1}2[1+(-1)^{\tau ^a_2 +\tau ^b_2}]\left[ (m^a_2n^b_2-n^a_2m^b_2)[1+ (-1)^{\tau ^a _1 +\tau ^b_1}] \right.+ \nonumber \\
&+& \left.(-1)^{\tau ^a_1+1}(n^a_2n^b_2+m^a_2m^b_2+m^a_2n^b_2)+(-1)^{\tau ^b_1}(n^a_2n^b_2+m^a_2m^b_2+n^a_2m^b_2) \right] \\
(56)(14) &\circ &  (56)(56) =0 
\eea
\subsection{$(n^{a}_1, m^{a}_1)=(n^{a}_3, m^{a}_3)=(1,0) \bmod 2$, \ $(n^{b}_1, m^{b}_1)=(n^{b}_3, m^{b}_3)=(1,1) \bmod 2$} \label{1011}
In this case $(i^a_1,i^a_2)=(56), \ (j^a_1,j^a_2)=(14)$ or $(56)$, and  $(i^b_1,i^b_2)=(45), \ (j^b_1,j^b_2)=(16)$ or $(45)$.
\bea
(56)(14) &\circ &  (45)(16) =(-1)^{\tau ^a_2}(56)(14)\circ (45)(45)= \nonumber \\
&=&(-1)^{\tau ^b_2} (56)(56) \circ (45)(45)=(-1)^{\tau ^a_2+\tau ^b_2}(56)(56) \circ (45)(16) =\nonumber \\
&=& (-1)^{\tau ^a _0 +\tau ^b _0 +1}2 \left[-(n^a_2n^b_2+m^a_2m^b_2+m^a_2n^b_2)[1+(-1)^{\tau ^a_1+\tau ^b_1}] \right. + \nonumber \\
&+&\left. (-1)^{\tau ^a_1}(n^a_2n^b_2+m^a_2m^b_2+n^a_2m^b_2)+(-1)^{\tau ^b_1+1}(n^a_2m^b_2-m^a_2n^b_2)           \right]  \label{ab5645}
\eea


\section{Computations}
Using the calculations presented in the previous  section and the appendices, we may compute the intersection numbers 
$a \circ b$ and $a \circ b'$ for any two stacks $a$ and $b$. 
We seek wrapping numbers $(n^{a,b}_k,m^{a,b}_k) \ (k=1,2,3)$ that determine the bulk parts $\Pi _{a,b}^{\rm bulk}$  and 
exceptional parts $\Pi _{a,b}^{\rm ex}$ of the two  supersymmetric stacks, that 
have no symmetric matter  and not too much antisymmetric matter on either stack, {\it i.e.} they
 satisfy the constraints (\ref{fphig}),(\ref{gAap6}),(\ref{fphigb}) 
and (\ref{gAbp6}), 
and that produce the required intersection numbers
\beq
(|a \circ b|,|a \circ b'|)=(\underline{1,2}) \label{abab121}
\eeq
On all lattices, it turns out that this is only possible when the wrapping numbers  of the two stacks are in {\em different} 
classes $ \bmod \ 2$, 
{\it i.e.} only the $(n^{a}_1, m^{a}_1)=(1,0) \bmod 2=(n^{a}_3, m^{a}_3)$, \  $(n^{b}_1, m^{b}_1)=(1,1) \bmod 2=(n^{b}_3, m^{b}_3)$ sector 
 can satisfy the constraints. It follows  that  the fixed points associated with $\Pi ^{\rm ex}_a$  and $\Pi ^{\rm ex}_b$ are
\bea
 (i^a_1,i^a_2)(j^a_1,j^a_2) &=&(14)(14), \ (14)(56), \ (56)(14), \ {\rm or} \ (56)(56) \\
 (i^b_1,i^b_2)(j^b_1,j^b_2) &=&(16)(16), \ (16)(45), \ (45)(16), \ {\rm or} \ (45)(45)
 \eea
As noted previously, it suffices to consider only the offset pairs of  fixed points  
$(i^a_1,i^a_2)=(56)$ and $(i^b_1,i^b_2)=(45)$ in $T^2_1$, since the results for the non-offset pairs may 
be obtained by setting $\tau^a_1=0$ and/or $\tau^b_1=0$. 
Further, we need only perform the computations for  the offset pairs  of 
fixed points $(j^a_1,j^a_2)=(56)$ and $(j^b_1,j^b_2)=(45)$ in $T^2_3$, 
since the results for $(a \circ b, a \circ b')$ in the other sectors
 are merely special cases that arise when $\tau ^a_2=0$ and/or $\tau ^b_2=0$. If, for example,
  $\tau ^a_2 = 1 \bmod 2$ is needed  in order to satisfy one 
 or more of the constraints, then we {\em must} use the offset fixed points $(j^a_1,j^a_2)=(56)$; otherwise, we may use either $(14)$ 
 or $(56)$. Similarly, if $\tau ^b_2= 1 \bmod 2$ then $(j^a_1,j^a_2)=(45)$ must be used; otherwise $(16)$ or $(45)$ may be used.
Thus,  in all cases the exceptional parts of the stacks are taken to be
\bea
 \Pi _a ^{\rm ex}=(56)(56)(n^a_2,m^a_2)=(-1)^{\tau^a _0}\{ [-(n^a_2+m^a_2)+(-1)^{\tau^a _1 }n^a_2][\epsilon _5+(-1)^{\tau^a _2}\epsilon _6 ] - \nonumber \\
-[n^a_2 +(-1)^{\tau^a _1}m^a_2][\tilde{\epsilon }_5+(-1)^{\tau^a _2}\tilde{\epsilon} _6 ] \}  \label{5656n2m2} \\
\Pi _b ^{\rm ex}=(45)(45)(n^b_2,m^b_2)=(-1)^{\tau^b _0}\{ [m^b_2+(-1)^{\tau ^b_1 +1}(n^b_2+m^b_2)][\epsilon _4+(-1)^{\tau^b _2}\epsilon _5 ] + \nonumber \\
+[n^b_2+m^b_2 +(-1)^{\tau^b _1 +1}n^b_2][\tilde{\epsilon }_4+(-1)^{\tau^b _2}\tilde{\epsilon} _5 ] \} \label{4545n2m2}
\eea
The orientifold duals  ${ \Pi _{a,b} ^{\rm ex}}'$ of course depend upon the lattice.

\subsection{AAB lattice}
For this lattice
\bea
 {\Pi _a ^{\rm ex}}'=(56)(56)(n^a_2,m^a_2)'=(-1)^{\tau^a _0}\{ [(n^a_2+m^a_2)-(-1)^{\tau^a _1 }n^a_2][\epsilon _6+(-1)^{\tau^a _2}\epsilon _5 ] + \nonumber \\
+[m^a_2 -(-1)^{\tau^a _1}(n^a_2+m^a_2)][\tilde{\epsilon }_6+(-1)^{\tau^a _2}\tilde{\epsilon} _5 ] \} \label{56561n2m2} \\
{\Pi _b ^{\rm ex}}'=(45)(45)(n^b_2,m^b_2)'=(-1)^{\tau^b _0}\{ [-m^b_2+(-1)^{\tau ^b_1 }(n^b_2+m^b_2)][\epsilon _4+(-1)^{\tau^b _2}\epsilon _6 ] + \nonumber \\
+[n^b_2 +(-1)^{\tau^b _1 }m^b_2][\tilde{\epsilon }_4+(-1)^{\tau^b _2}\tilde{\epsilon} _6 ] \} \label{45451n2m2}
\eea
On all lattices there are many sets of wrapping numbers for which the constraints (\ref{fphig}) and (\ref{fphigb}) 
that ensure the absence of 
symmetric matter on both stacks are satisfied. However, the vast majority of these do not have the required intersection numbers
 (\ref{abab121}). 
Here, and generally on the other lattices, the constraint (\ref{gAbp6}) that limits the amount of 
matter in antisymmetric representations of the gauge group, eliminates a lot of otherwise acceptable solutions. 
Consider, for example,  
\bea
(n^{a}_1, m^{a}_1)(n^a_2,m^a_2)(n^{a}_3, m^{a}_3)=(1,0)(1,0)(1,0)  \label{nama} \\
(n^{b}_1, m^{b}_1)(n^b_2,m^b_2)(n^{b}_3, m^{b}_3)=(1,1)(1,1)(1,-1) 
\label{nbmb}
\eea
 so that 
 \bea
 \Pi_a^{\rm bulk}&=&\rho_1  \label{a21} \\
 \Pi_b^{\rm bulk}&=&3(\rho _1+\rho _3-\rho _4-\rho _6)
 \eea
On the {\bf AAB} lattice these give
\bea
 {\Pi_a^{\rm bulk}}'&=&\rho_1 \\
 {\Pi_b^{\rm bulk}}'&=&-3\rho _6
 \eea 
Combining with the exceptional parts given in (\ref{5656n2m2}),(\ref{4545n2m2}),(\ref{56561n2m2}) and (\ref{45451n2m2}), we have 
\bea
a&=& \frac{1}{2}\rho _1+\frac{1}{2}(-1)^{\tau ^a_0}\{ [-1+(-1)^{\tau ^a _1}][{\epsilon}_5 +(-1)^{\tau ^a_2}{\epsilon}_6 ] 
- [\tilde{\epsilon}_5 +(-1)^{\tau ^a_2}\tilde{\epsilon}_6 ]  \}\\
a'&=& \frac{1}{2}\rho _1+\frac{1}{2}(-1)^{\tau ^a_0+\tau^a_1+\tau^a_2}\{ [-1+(-1)^{\tau ^a _1}][{\epsilon}_5 +(-1)^{\tau ^a_2}{\epsilon}_6 ] 
- [\tilde{\epsilon}_5 +(-1)^{\tau ^a_2}\tilde{\epsilon}_6 ] \} \\
b&=& \frac{3}{2}(\rho _1+\rho _3-\rho _4-\rho _6)+\frac{1}{2}(-1)^{\tau ^b_0}\{ [1+(-1)^{\tau ^b_1 +1}2][\epsilon _4 +(-1)^{\tau ^b_2}\epsilon _5]
+[2+(-1)^{\tau ^b_1 +1}] [\tilde{\epsilon}_4 +(-1)^{\tau ^b_2}\tilde{\epsilon}_5] \} \nonumber \\
&& \\
 b'&=&-\frac{3}{2}\rho _6+\frac{1}{2}(-1)^{\tau ^b_0}\{[-1+(-1)^{\tau ^b_1 }2][\epsilon _4 +(-1)^{\tau ^b_2}\epsilon _6]+
 [1+(-1)^{\tau ^b_1 }][\tilde{\epsilon}_4 +(-1)^{\tau ^b_2}\tilde{\epsilon}_6] \}
 \eea
Then both (\ref{fphig}) and, when $\tau ^b_1=0$,  (\ref{fphigb}) are satisfied, 
 so that there is no matter in the symmetric representation of either gauge group. 
In fact, we find that $g(A^a_p)$, defined in (\ref{gAp}), is zero, so that $a \circ \Pi _{\rm O6} =0$. 
As noted earlier, this means that there 
is no matter in the antisymmetric representation of the gauge group on the stack $a$ either.  
It is obvious from (\ref{nama}) that $a$ preserves supersymmetry,
  because the angles 
 $\phi ^a _k \ (k=1,2,3)$
that the 1-cycles with wrapping numbers $(n^a_k,m^a_k)$ on $T^2_k$ make with the Re $z_k$ axis are all 
 zero, independently of the the complex structure $U_3$. However, Im $U_3$
  {\em is} fixed when the  supersymmetry constraint $Y^b=0$, with $Y^b$ given in Table \ref{susy3cycle},
   is imposed on $b$. Since, from (\ref{tau3B}), the real part 
 Re $U_3= \frac{1}{2}$ we find
 \beq
   U_3=\frac{1}{2}+i\frac{\sqrt{3}}{2}=e^{i\pi /3}
 \eeq
Also  $X^b>0$ so that $b$ {\it is } supersymmetric, as required.  
In fact, from (\ref{nbmb}), $\phi^b_1=\phi^b_2=\pi/6, \ \phi^b_3=-\pi/3$. 
Thus $T^2_3$ has the same complex structure as  $T^2_{1,2}$,  given in (\ref{tau12}), and is therefore an $SU(3)$ lattice too.
 Finally, it is easy to verify from (\ref{aob})
 and (\ref{aob'}) using (\ref{ab5645}) and (\ref{5645ab'}) that
\bea
 a \circ b&=& \frac{3}{2} + \frac{1}{2}(-1)^{\tau^a_0+\tau^b_0+\tau ^b_2}[2-(-1)^{\tau ^a_1}] \\
a \circ b'&=&-\frac{3}{2} + \frac{1}{2}(-1)^{\tau^a_0+\tau^b_0 +\tau^a_2+\tau ^b_2}[1-2(-1)^{\tau ^a_1}] 
\eea
Consequently $(a \circ b, a \circ b') =(2,-1)$ or $(1,-2)$, provided that $\tau ^a_1=0$ and $\tau ^a_2=1 \bmod 2$.  
(It is often, but not invariably, the case that  we find two pairs of values of  $(a \circ b, a \circ b')$ 
both of which satisfy (\ref{abab121}).)
However, in this example $g(A^b_p)=12$, so that (\ref{gAbp6}) is {\em not} satisfied, and there are too many copies of
 the antisymmetric representation ${\bf A}_b$ 
to obtain the standard model matter content without vector-like quarks or leptons.

In fact, on this lattice we find just two independent pairs of wrapping number sets which, with a suitable choice of Wilson lines, 
satisfy all of the constraints. They are shown in Table \ref{aab}. 
\begin{table}
 \begin{center}
\begin{tabular}{||c|c||c|c||c||} \hline \hline
$(n^a_1,m^a_1;n^a_2,m^a_2;n^a_3,m^a_3)$&$(A^a_1,A^a_3,A^a_4,A^a_6)$&$(n^b_1,m^b_1;n^b_2,m^b_2;n^b_3,m^b_3)$&$(A^b_1,A^b_3,A^b_4,A^b_6)$ & Im $U_3$  \\ \hline \hline
$(1,0;1,0;1,0)$& $(1,0,0,0)$&$(1,-1;1,-1;-1,1)$& $(1,1,-1,-1)$ & $\frac{\sqrt{3}}{2}$ \\ \hline
$(1,-2;1,0;-1,2)$ &$(1,2,-2,-4)$ & $(1,-1;1,-1;-1,1)$& $(1,1,-1,-1)$& $\frac{\sqrt{3}}{2}$ \\ \hline \hline
\end{tabular}
\end{center} 
\caption{ \label{aab} Solutions on the {\bf AAB} lattice.}
 \end{table}
The first pair yields $(a \circ b, a \circ b') =(2,1)$ or  $(-1,-2)$, and the second pair $(-2,1)$ or  $(-1,2)$.
For both pairs there are sets of wrapping numbers that differ from those displayed by sign reversal of the pairs 
$(n^{a}_k, m^{a}_k)$ on two of the tori $T^2_k \ (k=1,2,3)$ and similarly for $(n^{b}_k, m^{b}_k)$,
and which also satisfy all of the constraints with a possibly different choice of Wilson lines. 
This is the case in all of our displayed solutions.
We  display just one representative in each set. In all cases, 
 $a \circ b $ and $a \circ b'$ is the same for all representatives.

As in the example discussed above, there is no matter in the antisymmetric representation 
${\bf A}_a$ on stack $a$ in either of our solutions. 
The two solutions have the same wrapping numbers on $b$ for which the number of antisymmetric representations  is
\beq
\# ({\bf A} _b) =\frac{1}{2}g(A^b_p) =2 \label{Abaab} 
\eeq
 
When $a \circ b$ and $a \circ b'$ have the same sign, as in the first pair,  $N_b=2$ 
  and ${\bf A} _b=\overline{\bf 1}$.  
Thus, in the first case there are no quark singlet states $q^c_L$ on $a$, only the $U(3)$ gauge particles; and
 there are  2 left-chiral lepton singlet states $\ell^c_L$ on the stack $b$, besides the $U(2)$ gauge particles. 
In contrast, in the second set where  $a \circ b$ and $a \circ b'$ have  opposite signs,  $N_b=3$ and ${\bf A} _b=\overline{\bf 3}$.  
Thus there are no lepton  singlet states $\ell ^c_L$ on $a$, only the $U(2)$ gauge particles; and there 
are  2 left-chiral quark singlet states $q^c_L$ on the stack $b$, besides the $U(3)$ gauge particles. 
As before, the supersymmetry 
constraint requires that the complex structure  $U_3=e^{i\pi /3}$, so that again $T^2_3$ is an $SU(3)$ lattice.
 As noted in the introduction, the antisymmetric representation carries $Q=2$ units of the relevant $U(1)$ charge. In 
the latter case, then, the 2  left-chiral quark singlet states $q^c_L$ contribute $Q_b=4$ units of $U(1)_b$ charge. 
However, the  3 quark doublets 
contribute $Q_b=6$ and the remaining 4 left-chiral quark singlets, arising from intersections of the stack $b$ with stacks $c,d,..$
each having  just one D6-brane, contribute $Q_b=-4$. Thus the overall cancellation of the charge $Q_b$ 
(required by RR tadpole cancellation) can {\em not} be 
achieved in this case. A similar argument applies in the former case when 2 left-chiral lepton singlet states arise in the 
antisymmetric representation on the $U(2)$ stack.

\subsection{BAB lattice}
For this lattice
\bea
 {\Pi _a ^{\rm ex}}'=(56)(56)(n^a_2,m^a_2)'=(-1)^{\tau^a _0}\{ [-m^a_2+(-1)^{\tau^a _1 }(n^a_2+m^a_2)][\epsilon _6+(-1)^{\tau^a _2}\epsilon _5 ] + \nonumber \\
+[n^a_2 +(-1)^{\tau^a _1}m^a_2][\tilde{\epsilon }_6+(-1)^{\tau^a _2}\tilde{\epsilon} _5 ] \} \label{bab56561n2m2} \\
{\Pi _b ^{\rm ex}}'=(45)(45)(n^b_2,m^b_2)'=(-1)^{\tau^b _0}\{ [-n^b_2+(-1)^{\tau ^b_1 +1}m^b_2][\epsilon _4+(-1)^{\tau^b _2}\epsilon _6 ] + \nonumber \\
-[n^b_2+m^b_2 +(-1)^{\tau^b _1 }n^b_2][\tilde{\epsilon }_4+(-1)^{\tau^b _2}\tilde{\epsilon} _6 ] \} \label{bab45451n2m2}
\eea
Here too we find  two independent pairs of wrapping number sets which, with a suitable choice of Wilson lines, 
satisfy all of the constraints. They are shown in Table \ref{bab}. 
\begin{table}
 \begin{center}
\begin{tabular}{||c|c||c|c||c||} \hline \hline
$(n^a_1,m^a_1;n^a_2,m^a_2;n^a_3,m^a_3)$&$(A^a_1,A^a_3,A^a_4,A^a_6)$&$(n^b_1,m^b_1;n^b_2,m^b_2;n^b_3,m^b_3)$&$(A^b_1,A^b_3,A^b_4,A^b_6)$ & Im $U_3$ \\ \hline \hline
$(1,0;1,-1;-1,2)$ &$(0,1,0,-2)$ & $(1,-1;1,-1;-1,1)$& $(1,1,-1,-1)$ & $\frac{1}{2\sqrt{3}}$\\ \hline 
$(1,-2;1,-1;-1,0)$ &$(2,1,0,0)$ &$(1,-1;1,-1;-1,1)$& $(1,1,-1,-1)$ & $\frac{1}{2\sqrt{3}}$\\ \hline \hline
\end{tabular}
\end{center} 
\caption{ \label{bab} Solutions on the {\bf BAB} lattice.}
 \end{table}
Both pairs have the property that there is no matter in the antisymmetric representation ${\bf A}_a$ on stack a, 
and that there are $\# ({\bf A} _b)=2$ on stack $b$, as in (\ref{Abaab}). 
Also,  $a \circ b$ and $a \circ b'$ have opposite signs for both pairs. 
 Thus in both solutions there are 2 quark singlet states $q ^c_L$ on $b$, besides the $U(3)$ gauge partices. 
 As for the {\bf AAB} lattice, overall cancellation of  $Q_b$ can {\em not} be achieved in this case either. 
 The imaginary part ${\rm Im} \ U_3$ 
 of the complex structure  on $T^2_3$ is fixed, as before, by the stack $b$. In neither case is an $SU(3)$  lattice required.

\subsection{ABB lattice}
As noted previously, for this lattice the orientifold duals ${\Pi _{a,b} ^{\rm ex}}'$ of the exceptional parts differ only by 
an overall sign from those on the {\bf BAB} lattice. Thus we merely reverse the signs of the right hand sides of (\ref{bab56561n2m2})
 and (\ref{bab45451n2m2}). Nevertheless different solutions do occur, as is apparent from the three solutions displayed in Table \ref{abb}.
\begin{table}
 \begin{center}
\begin{tabular}{||c|c||c|c||c||} \hline \hline
$(n^a_1,m^a_1;n^a_2,m^a_2;n^a_3,m^a_3)$&$(A^a_1,A^a_3,A^a_4,A^a_6)$&$(n^b_1,m^b_1;n^b_2,m^b_2;n^b_3,m^b_3)$&$(A^b_1,A^b_3,A^b_4,A^b_6)$& Im $U_3$ \\ \hline \hline
$(1,0;1,1;1,0)$& $(2,1,0,0)$&$(1,-1;0,1;1,-1)$& $(1,0,-1,0)$& $-\frac{1}{2\sqrt{3}}$\\ \hline
$(1,-2;1,1;1,-2)$ &$(0,-3,0,6)$ & $(1,-1;0,1;1,-1)$& $(1,0,-1,0)$& $-\frac{1}{2\sqrt{3}}$\\ \hline 
$(1,-2;1,-1;-1,0)$ &$(2,1,0,0)$ & $(1,-1;0,1;1,-1)$& $(1,0,-1,0)$&$-\frac{1}{2\sqrt{3}}$\\ \hline \hline
\end{tabular}
\end{center} 
\caption{ \label{abb} Solutions on the {\bf ABB} lattice.}
 \end{table}
In the first place, $(n^b_2,m^b_2)=(0,1) \bmod 2$ on this lattice, rather than $(1,1) \bmod 2$  on the {\bf BAB} lattice. 
This means that the exceptional part of $b$ is guaranteed to  be different on the two lattices. In any case, since the function 
$g(A^a_p)$ is different on the two lattices, the constraints that ensure the absence of antisymmetric matter differ, and we should 
expect different bulk solutions for $b$, as indeed we find. 

All three solutions share the property that, as before, there is no matter in the antisymmetric representation ${\bf A}_a$ on stack a. 
However, unlike all of the previous solutions, all of the solutions in Table \ref{abb}  also have the property that 
there is   no matter in the antisymmetric representation ${\bf A}_b$ on stack b. 
This might have been foreseen. The factor 12 that appears in the function $g(A^a_p)$  for the {\bf ABB} lattice in Table \ref{latfnfg}
requires that the only solutions that 
satisfy the constraints (\ref{gAap6}) and  (\ref{gAbp6}) must indeed have $g(A^a_p)=0=g(A^b_p)$, and hence have no antisymmetric 
(or symmetric) matter on $a$ or $b$.  A similar argument can be applied to the {\bf AAB} and {\bf  BAB} lattices in both of which 
the function $g(A^a_p)$ has a factor 4. Then satisfying the constraints (\ref{gAap6}) and  (\ref{gAbp6})
 requires that  $\#({\bf A}_{a,b}) \leq 2$, as we found. 
In all of the solutions displayed in Table \ref{abb},
 $a \circ b$ and $a \circ b'$ have opposite sign, so that $N_b=3$ and the gauge particles of 
$U(3)$ live on $b$. The complex structure $U_3$ of $T^2_3$ is fixed by the supersymmetry constraint on $b$, and, as we found  for the 
{\bf BAB} lattice,  an $SU(3)$ lattice is not required. The first and third pair of solutions have the same bulk parts for both 
$a$ and $b$. Nevertheless, they are distinct solutions since the exceptional parts of $a$ differ in the two pairs.

\subsection{BBB lattice}
For this lattice
\bea
 {\Pi _a ^{\rm ex}}'=(56)(56)(n^a_2,m^a_2)'=(-1)^{\tau^a _0}\{ [n^a_2+(-1)^{\tau^a _1 }m^a_2][\epsilon _6+(-1)^{\tau^a _2}\epsilon _5 ] + \nonumber \\
+[(n^a_2+m^a_2) -(-1)^{\tau^a _1}n^a_2][\tilde{\epsilon }_6+(-1)^{\tau^a _2}\tilde{\epsilon} _5 ] \} \label{bbb56561n2m2} \\
{\Pi _b ^{\rm ex}}'=(45)(45)(n^b_2,m^b_2)'=(-1)^{\tau^b _0}\{ [-(n^b_2+m^b_2)+(-1)^{\tau ^b_1 +1}n^b_2][\epsilon _4+(-1)^{\tau^b _2}\epsilon _6 ] + \nonumber \\
-[m^b_2 +(-1)^{\tau^b _1 +1}(n^b_2+m^b_2)][\tilde{\epsilon }_4+(-1)^{\tau^b _2}\tilde{\epsilon} _6 ] \} \label{bbb45451n2m2}
\eea
Since $(n^a_2,m^a_2)=(0,1) \bmod 2$   on this lattice only, the exceptional parts of any solutions 
 are guaranteed to differ from all previous solutions. In this case again we find three solutions. They are displayed in Table \ref{bbb}.
\begin{table}
 \begin{center}
\begin{tabular}{||c|c||c||c|c||c||} \hline \hline
$(n^a_1,m^a_1;n^a_2,m^a_2;n^a_3,m^a_3)$&$(A^a_1,A^a_3,A^a_4,A^a_6)$&$(n^b_1,m^b_1;n^b_2,m^b_2;n^b_3,m^b_3)$&$(A^b_1,A^b_3,A^b_4,A^b_6)$& Im $U_3$ \\ \hline \hline
$(1,0;2,-1;1,-2)$& $(1,-1,-2,2)$&$(1,-1;0,1;1,-1)$& $(1,0,-1,0)$& $-\frac{\sqrt{3}}{2}$\\ \hline
$(1,-2;2,-1;-1,0)$ &$(3,3,0,0)$ &$(1,-1;0,1;1,-1)$& $(1,0,-1,0)$& $-\frac{\sqrt{3}}{2}$\\ \hline 
$(1,-2;0,1;1,-2)$ &$(1,-1,-2,2)$ &$(1,-1;0,1;1,-1)$& $(1,0,-1,0)$& $-\frac{\sqrt{3}}{2}$\\ \hline \hline
\end{tabular}
\end{center} 
\caption{ \label{bbb} Solutions on the {\bf BBB} lattice.}
 \end{table}
The factor of 12 that appears in the function $g(A^a_p)$  in Table \ref{latfnfg} for this lattice too
 again ensures that any solution is guaranteed to 
have no antisymmetric (or symmetric) matter on either stack. Again 
$a \circ b$ and $a \circ b'$ have opposite sign in all solutions, so that $N_b=3$ and the gauge particles of 
$U(3)$ live on $b$. The supersymmetry constraint requires that  
the complex structure  $U_3= e^{-i\pi/3}$, 
so that $T^2_3$ is an $SU(3)$ lattice in this case. As for the {\bf ABB} lattice, the first and third pair of solutions 
are distinct since the exceptional parts of $a$ differ.
\section{Conclusions}
We have shown that, unlike the \Z$_6$ orientifold,  the   \Z$_6'$ orientifold 
{\em can} support  supersymmetric stacks $a$ and $b$ of D6-branes 
with intersection numbers satisfying $(|a \circ b|,|a \circ b'|)=(2,1)$ or $(1,2)$. Stacks having this property are an 
indispensable ingredient in any intersecting brane model that has {\em just} the matter content of the (supersymmetric) 
standard model. The number of branes, $N_{a,b}$ in stacks $a,b$ is required to be $(N_a,N_b)=(3,2)$ or $(2,3)$ so as to produce the 
gauge groups $U(3)$ and $U(2)$ from which the QCD $SU(3)_c$  and the electroweak $SU(2)_L$ gauge fields emerge.
 By construction, in all of our solutions  
there is no matter in symmetric representations of the gauge groups on either stack. However, some of the solutions {\em do} 
have matter, quark  singlets $q^c_L$ or lepton   singlets $\ell ^c_L$,  in  the antisymmetric representation  of gauge group
 on one of the stacks. This is not possible on the 
\Z$_6$ orientifold because all supersymmetric D6-branes wrap the same bulk 3-cycle as the O6-planes, from which it follows that 
$a \circ \Pi _{\rm O6}=0$. Then,  requiring the absence of symmetric matter necessarily entails the absence of antisymmetric 
matter too. In contrast, on the  \Z$_6'$ orientifold there exist supersymmetric 3-cycles that do not wrap the O6-planes. Thus, 
there is more latitude in this case, and the solutions with antisymmetric matter exploit this feature. 
 Unfortunately, however, none of the solutions of this nature that we have found can be enlarged to give just the 
 standard-model spectrum, since the overall cancellation of the relevant $U(1)$ charge cannot be achieved with just this matter content.  
Nevertheless, some of 
our solutions have no antisymmetric (or symmetric) matter on either stack. 
 We shall attempt in  a future work to construct  a  realistic (supersymmetric) standard model using one of our solutions.

The  presence of singlet matter on the branes in some, but not all,  of our solutions is an important feature of our results.
  It is clear that different orbifold point groups produce different physics, 
as indeed, for the reasons just given, our results also illustrate. The point group must act as an automorphism of the lattice used, but
 it is less clear that realising a given point group symmetry on 
different lattices produces different physics. Our results indicate that different lattices may produce different physics, 
since, for example, the solutions with no antisymmetric (or symmetric) matter on either stack occur only on the {\bf ABB} and {\bf BBB} 
lattices, and we understand why any acceptable solutions without symmetric matter must also lack antisymmetric matter. The 
observation that the lattice does affect the physics suggests that  other lattices are worth investigating in both the \Z$_6$ and 
\Z$_6'$ orientifolds. In particular, since $Z_6$ can be realised on a $G_2$ lattice, as well as  on an $SU(3)$ lattice, one or more of all 
 three $SU(3)$ lattices in the \Z$_6$ case, and of the two  on $T^2_{1,2}$ in the \Z$_6' $ case, could be replaced by a $G_2$ lattice. 
 We shall explore this avenue too in future work.
 
 The construction of a realistic model will, of course, entail adding further stacks of D6-branes $c,d,..$,  with just a single brane in 
 each stack, arranging that the matter content is just that of the supersymmetric standard model, the whole set 
 satisfying  (one of) the conditions (\ref{RRbulkaaa})...(\ref{RRbulkbbb}) and  the corresponding condition in 
 (\ref{RRexaaa})...(\ref{RRexbbb}) for RR tadpole cancellation.
 In a supersymmetric orientifold  RR tadpole cancellation ensures that NSNS tadpoles are also cancelled, 
  but some moduli, (some of)  of the complex structure moduli, the K\"ahler moduli 
 and the dilaton, remain unstabilised. Recent developments  have shown how such moduli may be stabilised using 
 RR, NSNS and metric fluxes \cite{Derendinger:2004jn,Kachru:2004jr,Grimm:2004ua,Villadoro:2005cu,DeWolfe:2005uu}, 
 and indeed C\'amara, Font \& Ib\'a\~nez \cite{Camara:2005dc, Aldazabal:2006up} have shown how  models 
 similar to the ones we have been discussing can be 
 uplifted into ones with stabilised K\"ahler moduli using a ``rigid corset''.
 In general,   such fluxes contribute to 
 tadpole cancellation conditions and might make them easier to satisfy. In contrast, the rigid corset 
 can be added to any RR tadpole-free assembly of D6-branes in order to stabilise all moduli. Thus our results represent 
 an important first step to obtaining a supersymmetric standard model from intersecting branes with all moduli stabilised.

\section{Acknowledgments}
 \nonumber
We are grateful to Gabriele Honecker for helpful correspondence and informative discussions, and especially for pointing out the 
error in the first version of this paper concerning
 the intersection numbers for exceptional cycles in the  $\theta^2$ and $\theta ^4$ twisted sectors. 
  We thank also the referee for pointing out that the intersection numbers $(|a \circ b|, |a \circ b'|)= (\underline{3,0})$
 allow the standard model spectrum when antisymmetric representations of $U(2)$ are present, 
 and also for suggesting several other helpful clarifications.
This work was funded in part by PPARC.

\appendix
\section{ Calculations of $(i^a_1,i^a_2)(j^a_1,j^a_2)\circ (i^b_1,i^b_2)(j^b_1,j^b_2)'$ on the {\bf AAB} lattice} \label{AAB}
\subsection{$(n^{a,b}_1, m^{a,b}_1)=(n^{a,b}_3, m^{a,b}_3)=(1,1) \bmod 2$}
As in \S \ref{11}, the fixed points involved are $(i^a_1,i^a_2),(i^b_1,i^b_2)=(56)$ and $(j^a_1,j^a_2),(j^b_1,j^b_2)=(14)$ or $(56)$
\bea
(45)(16) &\circ &  (45)(16)' = (45)(45) \circ (45)(45)' = \nonumber \\
&=& (-1)^{\tau ^a _0 +\tau ^b _0+1}2[m^a_2m^b_2+n^a_2m^b_2+m^a_2n^b_2 +(-1)^{\tau ^a_1+1}(n^a_2n^b_2+n^a_2m^b_2+m^a_2n^b_2) + \nonumber \\
&+&(-1)^{\tau ^b_1+1}(n^a_2n^b_2+m^a_2n^b_2+n^a_2m^b_2 )
+(-1)^{\tau ^a_1+\tau ^b_1}(n^a_2n^b_2-m^a_2m^b_2)]\\
(45)(16) &\circ &  (45)(45)' = (45)(45) \circ (45)(16)' = \nonumber \\
&=& (-1)^{\tau ^a _0 +\tau ^b _0 + \tau ^a_2+\tau ^b_2+1}2\left[n^a_2m^b_2+m^a_2n^b_2+m^a_2m^b_2  \right. + \nonumber \\
&+& \left.[(-1)^{\tau ^a_1+1}+(-1)^{\tau ^b_1 +1}](n^a_2n^b_2+n^a_2m^b_2+m^a_2n^b_2) 
 (-1)^{\tau ^a_1+\tau ^b_1}(n^a_2n^b_2-m^a_2m^b_2)\right]
\eea
\subsection{$(n^{a,b}_1,m^{a,b}_1)=(n^{a,b}_3,m^{a,b}_3)=(1,0)  \bmod 2$} \label{103}
In this case $(i^a_1,i^a_2),(i^b_1,i^b_2)=(56)$ and $(j^a_1,j^a_2),(j^b_1,j^b_2)=(14)$ or $(56)$, as in \S \ref{10}.
Consider first the exceptional brane
\bea
(56)(14)&=&(-1)^{\tau^a _0} \{   [-(n^a_2+ m^a_2)+(-1)^{\tau ^a _1}n^a_2] [\epsilon _1+(-1)^{\tau^a _2}\epsilon _4 ] -    \nonumber \\
 &-&   [n^a_2+(-1)^{\tau ^a _1}m^a_2] [\tilde{\epsilon }_1+(-1)^{\tau^a _2}\tilde{\epsilon} _4 ]   \}  \label{5614}
\eea
and its orientifold dual on the {\bf AAB} lattice
\bea
(56)(14)'&=&(-1)^{\tau^a _0} \{   [n^a_2+ m^a_2-(-1)^{\tau ^a _1}n^a_2] [\epsilon _1+(-1)^{\tau^a _2}\epsilon _4 ] -  \nonumber \\
 &-& [m^a_2-(-1)^{\tau ^a _1}(n^a_2+m^a_2)] [\tilde{\epsilon }_1+(-1)^{\tau^a _2}\tilde{\epsilon} _4]    \}  \label{5614'}
\eea
Since $(n^a_2,m^a_2)=(1,0) \bmod 2$
on this lattice
\bea
(56)(14)=(\tilde{\epsilon }_1+\tilde{\epsilon} _4 ) \bmod 2\\
(56)(14)'=(\tilde{\epsilon }_1+\tilde{\epsilon} _4 ) \bmod 2\\
\eea
Likewise
\bea
(56)(56)=(\tilde{\epsilon }_5+\tilde{\epsilon} _6 ) \bmod 2\\
(56)(56)'=(\tilde{\epsilon }_5+\tilde{\epsilon} _6 ) \bmod 2\\
\eea
It follows that for all of the allowed exceptional branes in this sector
\bea
(i^a_1,i^a_2)(j^a_1,j^a_2)\circ (i^b_1,i^b_2)(j^b_1,j^b_2)=0 \bmod 8 \\
(i^a_1,i^a_2)(j^a_1,j^a_2)\circ (i^b_1,i^b_2)(j^b_1,j^b_2)'=0 \bmod 8
\eea
The first of these may be verified from the  results in \S \ref{10}, using $(n^{a,b}_2,m^{a,b}_2)=(1,0) \bmod 2$.
Further,  $A^{a,b}_1=1 \bmod 2$ and $A^{a,b}_{3,4,6}=0 \bmod 2$ in this sector. In fact, since $A^{a,b}_1A^{a,b}_6=
 A^{a,b}_3A^{a,b}_4$, it follows that $A^{a,b}_6=0 \bmod 4$. Then $f_{AB}=0 \bmod 8=f_{AB'}$. 
Hence, from (\ref{aob}) and (\ref{aob'}), we see that 
 $a \circ b= 0 \bmod 2 = a \circ b'$ and we cannot obtain the required odd intersection number from this sector. We therefore omit the calculations of 
 $(i^a_1,i^a_2)(j^a_1,j^a_2)\circ (i^b_1,i^b_2)(j^b_1,j^b_2)'$ in this sector.
\subsection{$(n^{a}_1, m^{a}_1)=(n^{a}_3, m^{a}_3)=(1,0) \bmod 2$, \  $(n^{b}_1, m^{b}_1)=(n^{b}_3, m^{b}_3)=(1,1) \bmod 2$} \label{103113}
As in \S \ref{1011}, $(i^a_1,i^a_2)=(56), \ (j^a_1,j^a_2)=(14)$ or $(56)$, and  $(i^b_1,i^b_2)=(45), \ (j^b_1,j^b_2)=(16)$ or $(45)$.
\bea
(56)(14) &\circ &  (45)(16)' =(-1)^{\tau ^a_2}(56)(14)\circ (45)(45)'= \nonumber \\
&=&(-1)^{\tau^a_2+\tau ^b_2} (56)(56) \circ (45)(45)'=(-1)^{\tau ^b_2}(56)(56) \circ (45)(16)' =\nonumber \\
&=& (-1)^{\tau ^a _0 +\tau ^b _0 +1}2 \left[ -(n^a_2n^b_2+m^a_2n^b_2+n^a_2m^b_2)
+[(-1)^{\tau ^a_1}+(-1)^{\tau ^b_1}](n^a_2n^b_2-m^a_2m^b_2) \right.+ \nonumber \\
&+& \left.   (-1)^{\tau ^a_1+\tau ^b_1}(m^a_2m^b_2+m^a_2n^b_2+n^a_2m^b_2)                \right] 
\label{5645ab'}
\eea

\section{Calculations of $(i^a_1,i^a_2)(j^a_1,j^a_2)\circ (i^b_1,i^b_2)(j^b_1,j^b_2)'$ on the BAB lattice} \label{BAB}
\subsection{$(n^{a,b}_1,m^{a,b}_1)=(n^{a,b}_3,m^{a,b}_3)=(1,1)  \bmod 2$} 
As in \S \ref{11}, $(i^a_1,i^a_2),(i^b_1,i^b_2)=(45)$ and $(j^a_1,j^a_2),(j^b_1,j^b_2)=(16)$ or $(45)$. 
\bea 
(45)(16) &\circ &  (45)(16)' = (45)(45) \circ (45)(45)' = \nonumber \\
&=& (-1)^{\tau ^a _0 +\tau ^b _0}2\left[m^a_2m^b_2-n^a_2n^b_2 +[(-1)^{\tau ^a_1+1}+(-1)^{\tau ^b_1+1}](m^a_2m^b_2+n^a_2m^b_2+m^a_2n^b_2)\right. + \nonumber \\
&+& \left. (-1)^{\tau ^a_1+\tau ^b_1}(n^a_2n^b_2+n^a_2m^b_2+m^a_2n^b_2)\right]\\
(45)(16) &\circ &  (45)(45)' = (45)(45) \circ (45)(16)' = \nonumber \\
&=& (-1)^{\tau ^a _0 +\tau ^b _0 + \tau ^a_2+\tau ^b_2}2\left[m^a_2m^b_2-n^a_2n^b_2  \
[(-1)^{\tau ^a_1+1}+(-1)^{\tau ^b_1 +1}](m^a_2m^b_2+n^a_2m^b_2+m^a_2n^b_2) \right. + \nonumber \\
&+& \left. (-1)^{\tau ^a_1+\tau ^b_1}(n^a_2n^b_2+n^a_2m^b_2+m^a_2n^b_2)\right]
\eea
\subsection{$(n^{a,b}_1,m^{a,b}_1)=(n^{a,b}_3,m^{a,b}_3)=(1,0)  \bmod 2$}
As in \S \ref{10}, the relevant fixed points are $(i^a_1,i^a_2),(i^b_1,i^b_2)=(56)$ and $ (j^a_1,j^a_2),(j^b_1,j^b_2)=(14)$ or $(56)$.  
It follows from (\ref{5614}) that, since $(n^a_2,m^a_2)=(1,1) \bmod 2$
on this lattice
\beq
(56)(14)=({\epsilon }_1+{\epsilon} _4 )\bmod 2
\eeq
and, using Table \ref{Reps2}, that
\beq
(56)(14)'=({\epsilon }_1+{\epsilon} _4 ) \bmod 2
\eeq
Likewise
\bea
(56)(56)=({\epsilon }_5+{\epsilon} _6 ) \bmod 2\\
(56)(56)'=({\epsilon }_5+{\epsilon} _6 ) \bmod 2\\
\eea
 Thus, as in \S \ref{10}, for all of the allowed exceptional branes in this sector
\bea
(i^a_1,i^a_2)(j^a_1,j^a_2)\circ (i^b_1,i^b_2)(j^b_1,j^b_2)=0 \bmod 8 \\
(i^a_1,i^a_2)(j^a_1,j^a_2)\circ (i^b_1,i^b_2)(j^b_1,j^b_2)'=0 \bmod 8
\eea
Again, the first of these may be verified from the results of \S \ref{10} using  $(n^{a,b}_2,m^{a,b}_2)=(1,1) \bmod 2$. 
Further,  $A^{a,b}_3=1 \bmod 2$ 
and $A^{a,b}_{1,4,6}=0 \bmod 2$ 
in this sector. Since $A^{a,b}_1A^{a,b}_6=
A^{a,b}_3A^{a,b}_4$, it follows that $A^{a,b}_4=0 \bmod 4$, and again $f_{AB}=0 \bmod 8=f_{AB'}$.
 Thus,  from (\ref{aob}) and (\ref{aob'}), we see that 
 $a \circ b= 0 \bmod 2 = a \circ b'$ and, as before, we cannot obtain the required odd intersection number from this sector.
\subsection{$(n^{a}_1,m^{a}_1)=(n^{a}_3,m^{a}_3)=(1,0), \ 
(n^{b}_1,m^{b}_1)=(n^{b}_3,m^{b}_3)=(1,1) \bmod 2$} 
As in \S \ref{1011}, the relevant fixed points are  the fixed points involved are
 $(i^a_1,i^a_2)=(56), \ (j^a_1,j^a_2)=(14)$ or $(56)$, and  $(i^b_1,i^b_2)=(45), \ (j^b_1,j^b_2)=(16)$ or $(45)$.
\bea
(56)(14) &\circ &  (45)(16)' =(-1)^{\tau ^a_2}(56)(14)\circ (45)(45)'= \nonumber \\
&=&(-1)^{\tau^a_2+\tau ^b_2} (56)(56) \circ (45)(45)'=(-1)^{\tau ^b_2}(56)(56) \circ (45)(16)' =\nonumber \\
&=& (-1)^{\tau ^a _0 +\tau ^b _0 +1}2 \left[ m^a_2m^b_2+m^a_2n^b_2+n^a_2m^b_2)
+ (-1)^{\tau ^a_1+\tau ^b_1+1}(m^a_2m^b_2-n^a_2n^b_2)  \right.+ \nonumber \\
&+& \left.        [(-1)^{\tau ^a_1+1}+(-1)^{\tau ^b_1+1}](n^a_2n^b_2+n^a_2m^b_2+m^a_2n^b_2)         \right] 
\eea
\section{Calculations of $(i^a_1,i^a_2)(j^a_1,j^a_2)\circ (i^b_1,i^b_2)(j^b_1,j^b_2)'$ on the ABB lattice}
The fixed points involved are the 
same as for the {\bf BAB} (and the {\bf AAB}) lattice. 
Although the change in lattice  {\em does} affect the calculations of $(i^a_1,i^a_2)(j^a_1,j^a_2) \circ (i^b_1,i^b_2)(j^b_1,j^b_2)'$,
it does so only trivially. From Table \ref{Reps2}, we see that the only difference between the {\bf BAB} lattice and the {\bf ABB} lattice 
is that there is an overall minus sign in the orientifold image of the exceptional branes.
Thus, the results for this sector are trivially obtained by changing the overall sign  for the 
calculations of $(i^a_1,i^a_2)(j^a_1,j^a_2) \circ (i^b_1,i^b_2)(j^b_1,j^b_2)'$ in \S \ref{BAB}. Of course, this does not imply 
that the value $a \circ b'$ is also reversed in sign, since the contribution $f_{AB'}$ from the bulk branes is the same for 
both lattices. 
\section{Calculations of $(i^a_1,i^a_2)(j^a_1,j^a_2)\circ (i^b_1,i^b_2)(j^b_1,j^b_2)'$ on the BBB lattice}
\subsection{$(n^{a,b}_1,m^{a,b}_1)=(n^{a,b}_3,m^{a,b}_3)=(1,1)  \bmod 2$} 
As in \S \ref{11}, $(i^a_1,i^a_2),(i^b_1,i^b_2)=(45)$ and $(j^a_1,j^a_2),(j^b_1,j^b_2)=(16)$ or $(45)$. 

\bea
(45)(16) &\circ &  (45)(16)' = (45)(45) \circ (45)(45)' = \nonumber \\
&=& (-1)^{\tau ^a _0 +\tau ^b _0}2\left[ n^a_2n^b_2+n^a_2m^b_2+m^a_2n^b_2 +
[(-1)^{\tau ^a_1}+(-1)^{\tau ^b_1}](n^a_2n^b_2-m^a_2m^b_2) \right. + \nonumber \\
&+&\left. (-1)^{\tau ^a_1+\tau ^b_1}(m^a_2m^b_2+m^a_2n^b_2+n^a_2m^b_2)\right]\\
(45)(16) &\circ &  (45)(45)' = (45)(45) \circ (45)(16)' = \nonumber \\
&=& (-1)^{\tau ^a _0 +\tau ^b _0 + \tau ^a_2+\tau ^b_2}2\left[ n^a_2n^b_2+n^a_2m^b_2+m^a_2n^b_2 +
[(-1)^{\tau ^a_1}+(-1)^{\tau ^b_1}](n^a_2n^b_2-m^a_2m^b_2) \right. + \nonumber \\
&+&\left. (-1)^{\tau ^a_1+\tau ^b_1}(m^a_2m^b_2+m^a_2n^b_2+n^a_2m^b_2)\right]
\eea

\subsection{$(n^{a,b}_1,m^{a,b}_1)=(n^{a,b}_3,m^{a,b}_3)=(1,0)  \bmod 2$}
As in \S \ref{10}, the relevant fixed points are $(i^a_1,i^a_2),(i^b_1,i^b_2)=(56)$ and $(j^a_1,j^a_2),(j^b_1,j^b_2)=(14)$ or $(56)$.  
It follows from (\ref{5614}) and Table \ref{Reps2} that, since $(n^a_2,m^a_2)=(0,1) \bmod 2$
on this lattice
\bea
(56)(14)=({\epsilon }_1+{\epsilon} _4 +{\tilde{\epsilon }}_1+{\tilde{\epsilon}} _4) \bmod 2\\
(56)(14)'=({\epsilon }_1+{\epsilon} _4 +{\tilde{\epsilon }}_1+{\tilde{\epsilon}} _4) \bmod 2\\
\eea
Likewise
\bea 
(56)(56)=({\epsilon }_5+{\epsilon} _6 +{\tilde{\epsilon }}_5+{\tilde{\epsilon}} _6) \bmod 2\\
(56)(56)'=({\epsilon }_5+{\epsilon} _6+ {\tilde{\epsilon }}_5+{\tilde{\epsilon}} _6) \bmod 2\\
\eea
Thus, as in \S \ref{103}, for all of the allowed exceptional branes in this sector
\bea
(i^a_1,i^a_2)(j^a_1,j^a_2)\circ (i^b_1,i^b_2)(j^b_1,j^b_2)=0 \bmod 8 \\
(i^a_1,i^a_2)(j^a_1,j^a_2)\circ (i^b_1,i^b_2)(j^b_1,j^b_2)'=0 \bmod 8
\eea
However,  $A^{a,b}_{1,3}=1 \bmod 2$ and $A^{a,b}_{4,6}=0 \bmod 2$ in this sector, from which we 
can {\em not} conclude that either $f_{AB}$ or $f_{AB'}$ is  $0 \bmod 8$. 
In this case, therefore, we must compute 
$(i^a_1,i^a_2)(j^a_1,j^a_2)\circ (i^b_1,i^b_2),(j^b_1,j^b_2)'$.
\bea
(56)(14) &\circ &  (56)(14)' =(-1)^{\tau ^b_2} (56)(56) \circ (56)(56)' = \nonumber \\
&=& (-1)^{\tau ^a _0 +\tau ^b _0+1}2[1+(-1)^{\tau ^a_2 +\tau ^b_2}]\left[-(m^a_2m^b_2+n^a_2m^b_2+m^a_2n^b_2)  \right. + \nonumber \\
&+& \left. [(-1)^{\tau ^a_1}+(-1)^{\tau ^b_1}](n^a_2n^b_2+m^a_2n^b_2+n^a_2m^b_2 )
+(-1)^{\tau ^a_1+\tau ^b_1}(m^a_2m^b_2-n^a_2n^b_2)\right.]\\
(56)(14) &\circ &  (56)(56)' =0 =(56)(56) \circ (56)(14)'  
\eea

\subsection{$(n^{a}_1,m^{a}_1)=(n^{a}_3,m^{a}_3)=(1,0), \ 
(n^{b}_1,m^{b}_1)=(n^{b}_3,m^{b}_3)=(1,1)  \bmod 2$} 
As in \S \  \ref{1011}, the relevant fixed points are  the fixed points involved are
 $(i^a_1,i^a_2)=(56), \ (j^a_1,j^a_2)=(14)$ or $(56)$, and  $(i^b_1,i^b_2)=(45), \ (j^b_1,j^b_2)=(16)$ or $(45)$.
\bea
(56)(14) &\circ &  (45)(16)' =(-1)^{\tau ^a_2}(56)(14)\circ (45)(45)'= \nonumber \\
&=&(-1)^{\tau^a_2+\tau ^b_2} (56)(56) \circ (45)(45)'=(-1)^{\tau ^b_2}(56)(56) \circ (45)(16)' =\nonumber \\
&=& (-1)^{\tau ^a _0 +\tau ^b _0 +1}2 \left[ n^a_2n^b_2-m^a_2m^b_2
+[(-1)^{\tau ^a_1}+(-1)^{\tau ^b_1}](m^a_2m^b_2+m^a_2n^b_2+n^a_2m^b_2) \right.+ \nonumber \\
&+& \left.   (-1)^{\tau ^a_1+\tau ^b_1+1}(n^a_2n^b_2+m^a_2n^b_2+n^a_2m^b_2)                \right] 
\eea

\end{document}